\documentclass [prl,twocolumn,notitlepage,superscriptaddress,floatfix,bibliography] {revtex4-1}
\usepackage{amssymb}
\usepackage{setspace}
\usepackage{amsmath}
\usepackage{amsfonts}
\usepackage{hyperref}
\usepackage{graphicx}
\usepackage{float}
\usepackage{subfigure}
\usepackage{color}

\newcommand{\vect}[1]{\boldsymbol{#1}}
\newcommand{\hsa}{h_{\text{sat}}}
\newcommand{\Zbb}{\mathbb{Z}}
\newcommand{\Vbar}{\bar{\text{V}}}

\newcommand{\Hcal}{\mathcal{H}}

\newcommand{\mc}[1]{\mathcal{#1}}

\newcommand{\cond}[1]{\langle #1 \rangle}
\newcommand{\ad}{a^{\dagger}}
\newcommand{\bd}{b^{\dagger}}
\newcommand{\cd}{c^{\dagger}}
\newcommand{\dd}{d^{\dagger}}

\newcommand{\phid}{\phi^{\dagger}}
\newcommand{\ud}{u^{\dagger}}
\newcommand{\vd}{v^{\dagger}}
\newcommand{\ed}{e^{\dagger}}
\newcommand{\fd}{f^{\dagger}}
\newcommand{\psid}{\psi^{\dagger}}

\newcommand{\ta}{\tilde{a}}
\newcommand{\tad}{\tilde{a}^{\dagger}}
\newcommand{\tb}{\tilde{b}}

\newcommand{\tc}{\tilde{c}}

\newcommand{\td}{\tilde{d}}
\newcommand{\tdd}{\tilde{d}^{\dagger}}

\newcommand{\kv}{\vect{k}}

\newcommand{\qv}{\vect{q}}

\newcommand{\be}{\begin{equation}}
\newcommand{\ee}{\end{equation}}
\newcommand{\ba}{\begin{align}}
\newcommand{\ea}{\end{align}}
\newcommand{\non}{\nonumber}

\newcommand{\phik}{\varphi_{\kv}}
\newcommand{\thk}{\theta_{\kv}}
\newcommand{\tthk}{\tilde\theta_{\kv}}
\newcommand{\tthq}{\tilde\theta_{\qv}}
\allowdisplaybreaks

\begin{document}
\title{Half-magnetization plateau in a Heisenberg antiferromagnet  on a triangular lattice}
\author{Mengxing Ye}
\author{Andrey V. Chubukov}
\affiliation{School of Physics and Astronomy, University of Minnesota, Minneapolis, MN 55455, USA}

\date{\today}
\begin{abstract}
We present the phase diagram in a magnetic field of a 2D isotropic Heisenberg antiferromagnet on a triangular lattice. We consider  spin-$S$ model with nearest-neighbor ($J_1$) and next-nearest-neighbor ($J_2$) interactions. We focus on  the range of $1/8<J_2/J_1<1$, where the ordered states are different
from those in the model with only nearest neighbor exchange. A classical ground state in this range has four sublattices and is infinitely degenerate in any field. The  actual order is then determined by quantum fluctuations via ``order from disorder" phenomenon.  We argue that the phase diagram is rich due to competition between  competing four-sublattice quantum states which break either $\Zbb_3$ orientational symmetry or $\Zbb_4$ sublattice symmetry.
 At small and high fields, the ground state is a  $\Zbb_3$-breaking canted stripe state, but at intermediate fields  the ordered states break $\Zbb_4$ sublattice symmetry. The most noticeable of such states is ``three up, one down"  state in which spins in three sublattices are directed along the field and in one sublattice opposite to the field. Such a state breaks  no continuous symmetry and has gapped excitations. As the consequence, magnetization
has a plateau at exactly one half of the saturation value. We  identify gapless states, which border the ``three up, one down" state and discuss the transitions between these states and the canted stripe state.
\end{abstract}
\maketitle
\textbf{Introduction\quad}
Recent experimental and theoretical advances renewed the interest in the physics of  frustrated spin systems. In many of these systems the classical ground state is infinitely degenerate, and the actual ground state spin configuration is selected by quantum fluctuations (the ``order from disorder" phenomenon). The resulting ground state is often rather unconventional and in several cases displays a non-monotonic behavior of magnetization in an applied field, with kinks, jumps, and plateaus~\cite{Coldea2002,Tokiwa2006,Takigawa2011,Starykh2015,Wosnitza2016,Chubukov2014,Yamamoto2017}. The most known example of such behavior is in the case of a two-dimensional (2D) quantum antiferromagnet on a triangular lattice with nearest-neighbor exchange $J_1$~\cite{Chubukov1991,Mila2016}. Classically, all spin configurations, which satisfy $\vect{S}_{r}+\vect{S}_{r+\vect{\delta}_1}+\vect{S}_{r+\vect{\delta}_2}=\vect{h}S/(3J_1)$ for each triad of neighboring spins, have the same ground state energy. Quantum fluctuations lift the degeneracy and select a set of three coplanar configurations, between which the systems transforms upon increasing field. The middle configuration, which exists at $h$ around $1/3$ of the saturation field $h_{sat} =9J_1$ ($h$ scaled), is a collinear state with two spins up (U) and one spin down (D) in every elementary triangle (an UUD state). In such a state only a discrete $\Zbb_3$ symmetry is broken (one spin in a triad is selected to be antiparallel to a field), and, as a result, all excitations are gapped and the magnetization has a plateau at exactly one-third of the saturation value~\cite{Chubukov1991,Nikuni1993,Griset2011,Balents2013,Mila2016}. This plateau has been observed in Cs$_2$CuBr$_4$~\cite{Ono2003,Honecker2004,Ono2005,Tsujii2007,Fortune2009} and in Ba$_3$CoSb$_2$O$_9$~\cite{Susuki2013}. An UUD state survives in a finite range of perturbations, like the spatial anisotropy of the exchange interaction~\cite{Nikuni1993,Alicea2005,Alicea2009,Motrunich2010,Griset2011,Balents2013,Chubukov2013}, multiple-spin ring exchange~\cite{Kubo1997},
and the next nearest neighbor exchange~\cite{Ye2017a}, as long as the perturbations are not strong enough to close the minimal excitation gap in the UUD state. What replaces the UUD state at larger perturbations has been the subject of intensive research over the last several years~\cite{Alicea2009,Motrunich2010,Balents2013,Chubukov2013,Mila2013,Chalker2005,Chalker2006,Nikuni1993,Ueda2009,Griset2011}

\begin{figure}[htb!]
  \centering
  \includegraphics[width=1\linewidth]{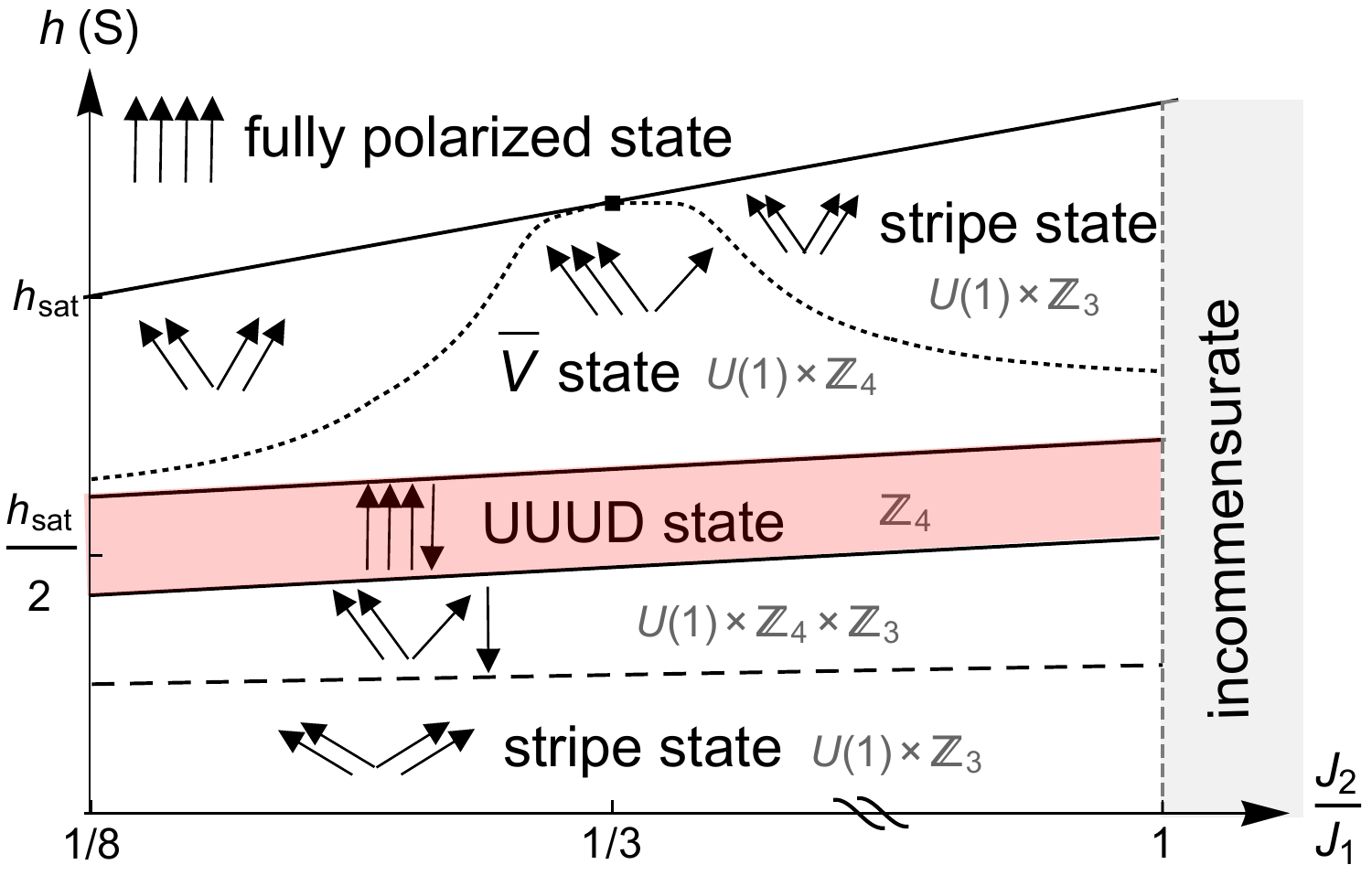}
  \caption{Semiclassical phase diagram of a spin-$S$, $J_1-J_2$ antiferromagnet on a triangular lattice,  at $1/8<J_2/J_1<1$. Solid (dotted) lines are second-order (first-order) phase transitions, which we identified and analyzed in this work. Dashed line is a first-order transition, which we expect to hold, but didn't analyze.   Arrows indicate magnetic order in the four-sublattice representation, and symbols like $U(1) \times Z_3$ indicate the broken symmetry in each state.
 \label{fig:PD}}
\end{figure}

\begin{figure}[htb!]
  \centering
  \includegraphics[width=0.9\linewidth]{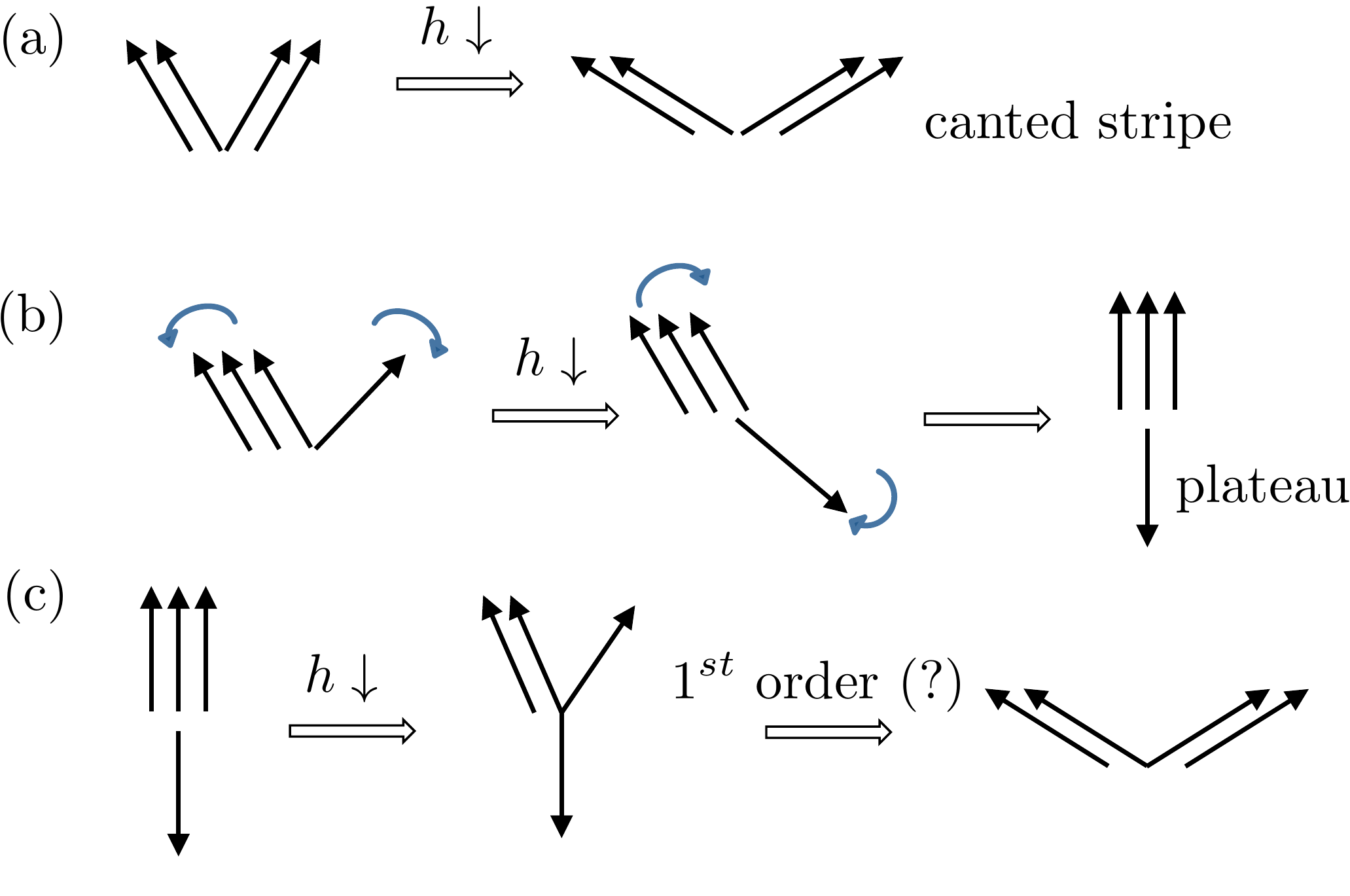}
  \caption{(a), (b) -- two candidate quantum  four-sublattice ground states upon decreasing of the magnetic field $h$ towards a half of saturation value. (a) A $Z_3$ breaking canted stripe state. As field goes down, the angle between two pairs of parallel spins increases. (b) $\Vbar$  and UUUD states. Both break $Z_4$ sublattice symmetry by selecting one sublattice with a different spin orientation compared to the other three. (c) Evolution from the UUUD state  to the canted stripe state as $h$ decreases below $h_{sat}/2$. \label{fig:spinOD}}
\end{figure}

\begin{figure}[htb!]
  \centering
  \subfigure[]{\includegraphics[width=0.4\linewidth]{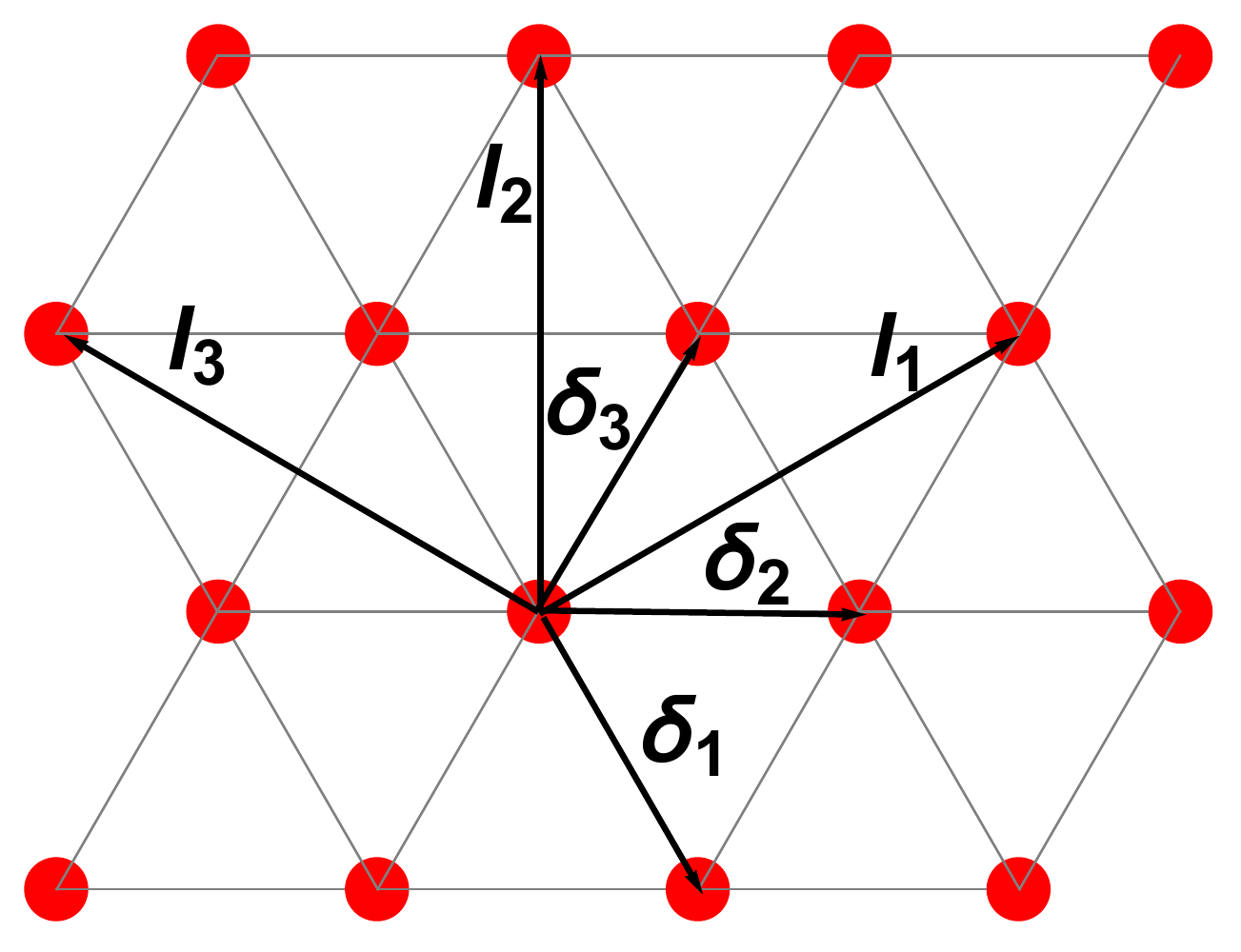}\label{fig:lattice}}\qquad
  \subfigure[]{\includegraphics[width=0.4\linewidth]{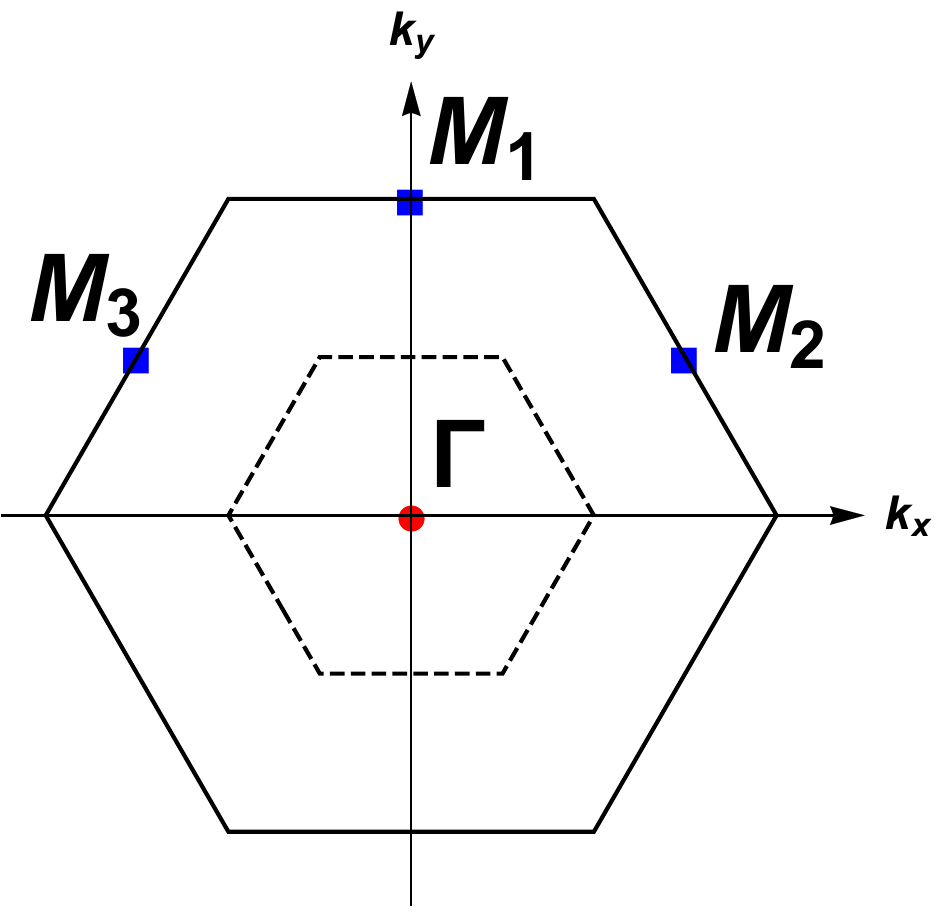}\label{fig:BZ}}
  \caption{(a) The nearest-neighbor ($\vect{\delta}_{i}$) and next-nearest-neighbor ($\vect{l}_{i}$)  bonds on a triangular lattice.
  (b) Solid (dashed) line: Single-sublattice (four-sublattice) Brillouin zone. The points labeled as $M_i$ are relevant to our discussion of spin-wave excitations near $\hsa$. Whereas the spin-wave excitations of the four-sublattice UUUD state soften at $\Gamma$ point.
  }
\end{figure}

 In this communication we study  spin-$S$  Heisenberg antiferronmagnet on a triangular lattice with nearest ($J_1$) and second-nearest ($J_2$) exchange interaction. Previous studies have found that the UUD phase and other two three-sublattice coplanar ground states in a field are immune to $J_2$ up to $J_2/J_1<1/8$. At larger $J_2$, however, the set of classical ground states changes discontinuously from three-sublattice configurations to four sublattice ones, in which four spins on two neighboring triads satisfy $\vect{S}_{r}+\vect{S}_{r+\vect{\delta}_1}+\vect{S}_{r+\vect{\delta}_2}+\vect{S}_{r+\vect{\delta}_3}=\vect{h}S/(2(J_1+J_2))$ (see Fig.~\ref{fig:lattice}). This condition does not uniquely specify spin order, even at zero field.  The  selection of the order by quantum fluctuations at $h=0$ has been analyzed  by various means~\cite{Jolicoeur1990,Campbell2015, Sheng2015,White2015}, and the consensus is that for $1/8<J_2/J_1<1$ the winner is the stripe order with ferromagnetic alinement of spins along one of three principle axes on a triangular lattice and antiferromagnetic along the other two.
  The same order (the canted stripe state, see Fig.~\ref{fig:spinOD}(a)) is selected by quantum fluctuations near the saturation field, and semiclassical (large $S$) spin-wave analysis shows~\cite{Ye2017a} that this  state remains stable at all fields. It would seem natural to conjecture that this state, with monotonic magnetization $M(h)$, is the true quantum ground state for $1/8<J_2/J_1<1$ in all fields.

\begin{figure}[htb!]
  \centering
  \includegraphics[width=0.8\linewidth]{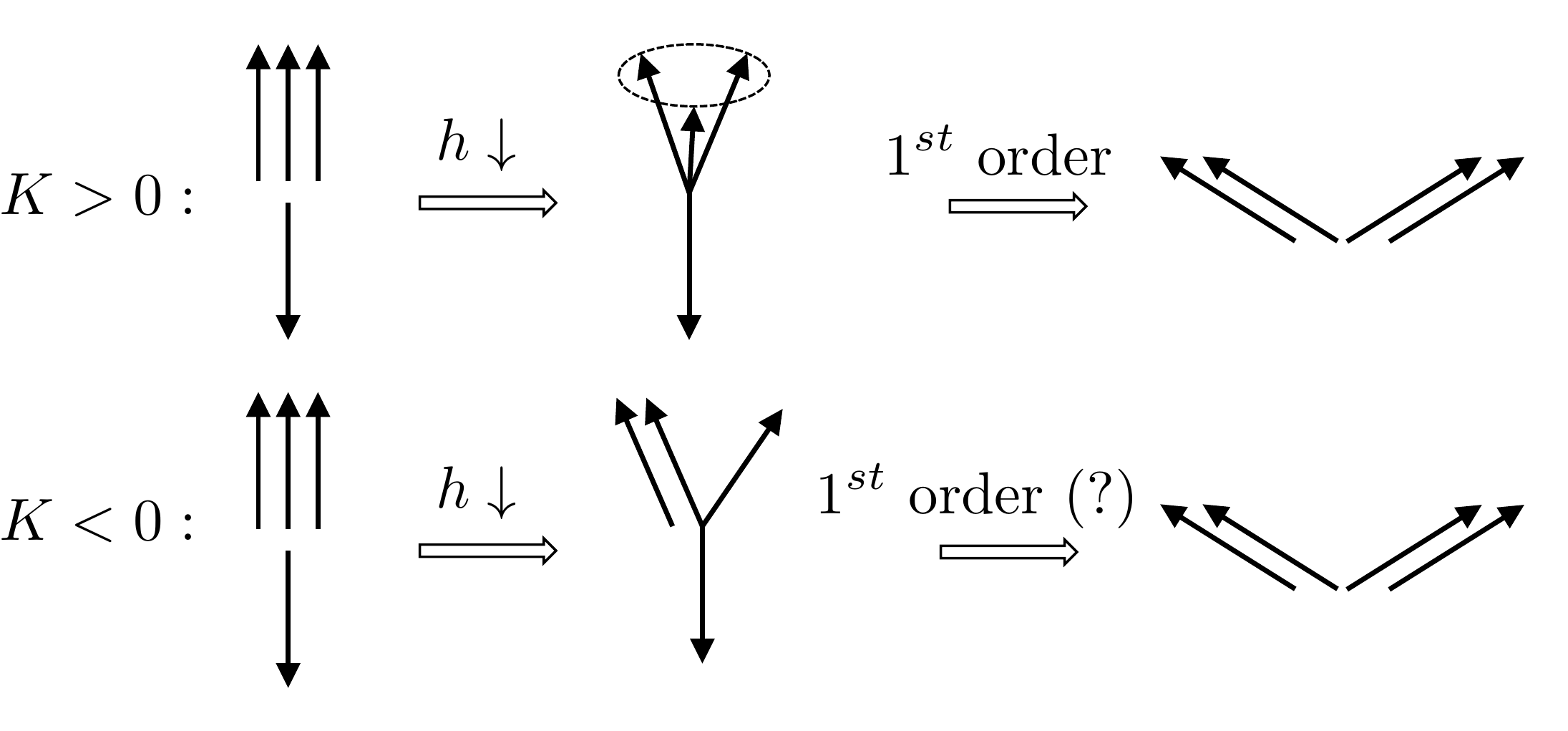}
  \caption{Evolution of the magnetic order below the UUUD state, depending on sign of the $K$  term in Eq.~\ref{eq:QuarticFE}.
  \label{fig:plateauDown}
  }
\end{figure}

We argue that the phase diagram of $J_1-J_2$ model in a field is actually rather complex, with multiple phases (see Fig.~\ref{fig:PD}), and the stripe order is the ground state configuration only in some range of fields and of $J_2/J_1$. For other values of $h$ and $J_2/J_1$ the ground state configurations are the co-planar states, similar to those at small $J_2$. In particular, around $ h = h_{sat}/2$, the ground state is the UUUD state, in which spins in three sublattices are aligned along the field and in the forth sublattice opposite to the field.  This spin order breaks $\Zbb_4$ sublattice symmetry, but doesn't break any continuous symmetry.  As a result, spin-wave excitations are gapped, and the magnetization has a plateau at exactly $1/2$ of the saturation value. We argue that the UUUD state exists for all $J_2$ in the interval $1/8<J_2/J_1<1$, i.e., the magnetization plateau exists for all $J_1-J_2$ systems, either at $1/3$ of the saturation value, at $J_2/J_1 < 1/8$, or at $1/2$ of the saturation value, at $1/8<J_2/J_1<1$.   We also  analyze the proximate states to the UUUD state. Above the upper critical field $h_u$, the UUUD state becomes unstable towards a state in which three up-spins rotate in one direction from the direction of $\vect{h}$, and  the down-spin rotates in the opposite direction (see Fig.~\ref{fig:spinOD}(b)). Below the lower critical field $h_l$, we found, at large $S$, a particular coplanar state, in which down-spin does not move, while three up-spins again rotate, but now one of these three spins splits from the other two (see Fig.~\ref{fig:spinOD}(c)). A non-coplanar, chiral umbrella state~\cite{Korshunov1993,Kubo1997} is close in energy and may be the ground state near $h_l$ at smaller $S$ (see Fig.~\ref{fig:plateauDown}).

A cascade of field-induced magnetic transitions at fields below $h_{sat}/2$ has been observed in 2$H-$AgNiO$_2$~\cite{Coldea2014,Coldea2008}. It has been argued~\cite{Wheeler2009} that in this material Ni$^{2+}$ ions are localized and form a $S=1$ triangular lattice antiferromagnet with $J_2 = 0.15 J_1$, 
single-ion easy axis anisotropy $D$, weak ferromagnetic exchange between layers. And  Classical Monte-Carlo calculations for this model have found~\cite{Seabra2010} the region of UUUD phase, whose width at $T=0$ scales with $D$. We show that in a quantum model the UUUD phase is stable in a finite range of $h$ already at $D=0$. We expect that future measurements of the magnetization in 2$H-$AgNiO$_2$ at higher fields will be able to detect the UUUD phase and also the cascade of phases above $h_{sat}/2$. The analysis of the high-field phases will allow one to distinguish whether UUUD order is stabilized predominantly by quantum fluctuations or by single-ion anisotropy~\footnote{If the UUUD order is dominated by quantum fluctuations, one should expect to see both ${\bar V}$ phase and canted stripe phase at higher fields, like in Fig.~\ref{fig:PD}. If UUUD order is mostly due to single-ion anisotropy, only ${\bar V}$ phase is present, see Ref.~\cite{Seabra2011}.}
 %
%

\textbf{Model and the high field phase diagram\quad}
The $J_1-J_2$ Heisenberg antiferromagnet on a triangular lattice is described by
\begin{align}
\mathcal{{ H}}&={ J}_1\sum_{\langle i,j\rangle}\vect{S}_i\cdot\vect{S}_j+{ J}_2\sum_{\langle\langle i,j\rangle\rangle}\vect{S}_i\cdot\vect{S}_j-S\vect{{
h}}\cdot\sum_{i} \vect{S}_i
\label{eq:modelH}
\end{align}
where $\langle i , j\rangle$ and $\langle\langle i , j \rangle\rangle$ run over all the nearest and next nearest neighbor bonds. Due to the global spin-rotational symmetry, the direction of $\vect{h}$ does not matter. We choose $\vect{h}= h \,\hat{e}_z$, and consider the range of $1/8<J_2/J_1<1$.

The first indication that the stripe phase is not the only ground state in a field comes from the Ginzburg-Landau analysis of the order immediately below the saturation field. We said in the Introduction that this analysis yields the stripe order. This is true for all $J_2$ in the interval of interest, however, with one exception --  $J_2=J_1/3$.  To see why this $J_2$ is exceptional, we note that spin-wave excitations soften at $h=h_{sat}$ at three points in the Brillouin zone ($M_1$, $M_2$, $M_3$ in Fig.~\ref{fig:BZ}). To understand the order below $h_{sat}$ one then needs to introduce three condensates  $\Phi_1,\,\Phi_2,\,\Phi_3$. The ground state energy in terms of  $\Phi$ is:
\begin{align}\label{eq:Eden11}
&E_{\Phi}/N=
-\mu\sum_{i=1,2,3}|\Phi_i|^2+\frac{1}{2}\Gamma_1\sum_{i=1,2,3}|\Phi_i|^4\non\\&+
\Gamma_2(|\Phi_1|^2|\Phi_2|^2+|\Phi_1|^2|\Phi_3|^2+|\Phi_2|^2|\Phi_3|^2)\non\\&+
\Gamma_3(\Phi_1^2\Phi_2^2+\Phi_2^2\Phi_3^2+\Phi_3^2\Phi_1^2+h.c.)
\end{align}
where $\mu\sim S(\hsa-h)$. The type of spin order that  minimizes $E_{\Phi}$   depends on the interplay between the quartic coefficients $\Gamma_i$. In the classical limit, $\Gamma_1=\Gamma_2=8(J_1+J_2),~\Gamma_3=0$, i.e., any state from the manifold $|\Phi|_1^2+|\Phi|_2^2+|\Phi|_3^2\equiv \mu/\Gamma_1$ is the ground state. Quantum fluctuations lift the degeneracy. To leading order in $1/S$ we found~\cite{Ye2017a}, near $J_2=J_1/3$,
\begin{align}
\Gamma_2-\Gamma_1 &= \frac{24\sqrt{3}J_1}{\pi}~\big(\frac{J_2}{J_1}-1/3\big)^2 \frac{|\log (\hsa-h)|}{S}-\beta_1/S\non\\
 \Gamma_3&=-\beta_2/S
\label{eq:dGamma}
\end{align}
where $\beta_{1,2}>0$ are numbers of order one.  The logarithm $|\log (\hsa-h)|$ is present because of quadratic dispersion near $M$-points  in Fig.~\ref{fig:BZ}:
e.g., near ${\bf M}_1$,  $\omega_{\kv}=SJ_1((1+\frac{9}{2}\alpha)k_x^2+(1-\frac{3}{2}\alpha)k_y^2)-\mu$, where $\qv=\kv+\vect{M}_1$ and $\alpha=J_2/J_1-1/3$.
Because of the logarithm, $\Gamma_2 > \Gamma_1$, A straightforward analysis then shows that  only one $\Phi_i$ is non-zero because it costs extra energy to develop simultaneously  condensates from different valleys.  The resulting order is the stripe state. A selection of $\Phi_i$ breaks $\Zbb_3$ symmetry, which  for the stripe state can be understood as an orientational symmetry (spins align ferromagnetically along one of the three spatial directions).  However, the prefactor for the logarithm in $\Gamma_2-\Gamma_1$ in Eq.~\ref{eq:dGamma} is non-zero only when the dispersion is anisotropic, and it vanishes at $J_2=J_1/3$, when $\omega_{\kv}$ becomes isotropic ($\alpha=0$). For this $J_2/J_1$, the sign of
$\Gamma_2-\Gamma_1$ is determined by regular $1/S$ terms, along with the sign of $\Gamma_3$.  We computed these terms and found $\Gamma_2-\Gamma_1<0$, $\Gamma_3<0$. As a result, at $J_2/J_1=1/3$, all three condensates emerge with equal amplitudes and relative phases  0 or $\pi$ (because $\Gamma_3<0$). The four choices for
 $(\Phi_1, \Phi_2, \Phi_3)$ are $(\Phi, \Phi,  \Phi),~( \Phi, -\Phi, -\Phi),~(-\Phi,  \Phi, -\Phi),~(-\Phi, -\Phi,  \Phi)$. In each of these states spins in three sublattices tilt to one direction from the field, and in one sublattice tilt to the opposite (see Fig.~\ref{fig:PD}). We label such a state $\Vbar$ by analogy with the corresponding $V$ state \footnote{The three-sublattice $V$ state has spins in two sublattices tilt in one direction from the field, and in another sublattice to the opposite direction.} at $J_2 < J_1/8$~\cite{Nikuni1993,Griset2011,Balents2013,Mila2016}. The $\Vbar$ state breaks $U(1)$ spin-rotational symmetry in the plane perpendicular to the field, and also breaks a $\Zbb_4$ sublattice symmetry by selecting a sublattice in which spin direction is different from that in other three sublattices.

Immediately below $h_{sat}$, the $\Vbar$ state is stable in the infinitesimally small range around $J_2=J_1/3$, at $(J_2/J_1-1/3)^2 < 1/|\log(h_{sat}-h)|$. As $h$ decreases, the width grows and  becomes $\mathcal{O}(1)$ at   $\hsa-h=\mathcal{O}(1)$.  The $\Vbar$ and the stripe state break different discrete symmetries ($\Zbb_4$ and $\Zbb_3$, respectively), hence the transition between the two states is likely first order. The increase of the width of the $\Vbar$ state with decreasing field can be understood as a generic consequence of the fact that  this state is favored by regular $1/S$ terms, i.e., by quantum  fluctuations at short length scales, while the stripe phase   is favored by $|\log (\hsa-h)|$, which comes from  long-wavelength fluctuations. As the magnitude of the transverse order increases  with decreasing field,
long wavelength fluctuations are suppressed, and  $\Vbar$ state becomes more favorable.

\textbf{Half-magnetization plateau\quad}
As the field decreases towards $h_{sat}/2$, the  $\Vbar$ state evolves: the spin in one sublattice continuously rotates away from the field direction towards the direction antiparallel to $\vect{h}$. The spins in three other sublattices remain parallel to each other and first rotate away from the field, and then rotate back. Eventually, near $h = h_{sat}/2$, spins in the three sublattices become parallel to $\vect{h}$ and spins in the fourth sublattice become antiparallel to $\vect{h}$ (see Fig.~\ref{fig:spinOD}(b)). Once this happens, the system enters into the new, UUUD phase.  In this phase, $U(1)$ symmetry is restored (there is no sublattice spin component transverse to the field), but $Z_4$ symmetry is still broken.  To obtain the boundaries of the UUUD phase, we compute its excitation spectrum. For this, we introduce four sets of Holstein-Primakoff (H-P) bosons and do spin-wave calculations to order $1/S$.    In the classical, $S \to \infty$ limit, the spin-wave excitations are stable only at $h = h_{sat}/2$, where the spectrum consists of one gapped spin wave branch (in-phase precession of all spins around the field), and three gapless branches, with zero modes at $\Gamma$ point of the four-sublattice Brillouin zone (see Fig.~\ref{fig:BZ}). Quantum $1/S$ correction to spectrum, however, make it stable in a finite range of $h$ around $h_{sat}/2$. Namely, all spin-wave branches become gapped (and positive) in a range $h_l<h<h_u$, where $h_l=h_{sat}/2-\delta_1$ and $h_u=h_{sat}/2+\delta_2$. We show the details of the calculations in the Supplementary Material (SM) and present the results for $\delta_1$ and $\delta_2$ in Table~\ref{tab:spectrumcorr}. We found, somewhat unexpected, that the stability width of the UUUD phase is finite for {\it all} $J_2$ in the interval $1/8 < J_2/J_1 <1$.  We further computed the ground state energy of the UUUD phase to order $1/S$ (classical energy plus $1/S$ corrections from zero point fluctuations), and compared with that of the stripe phase. We found that for all $J_2$ the energy of the UUUD state is lower.  Because of this and because the UUUD state naturally emerges from the $\Vbar$ state, we argue that the UUUD state is the true ground state near $h = h_{sat}/2$ for all $1/8 < J_2/J_1 <1$. As all excitations in the UUUD state are gapped, this state has magnetization fixed at exactly $1/2$ of the saturation value.

We also verified that at the upper critical field of the UUUD state, it becomes unstable towards $\Vbar$ state. Namely, at $h = h_u$ one of the spin-wave branches condenses, and the condensate leads to $\langle S_x\rangle =a$ for spins on three up-spin sublattices, and $-3a$ for the spins on the down-spin sublattice.  This result in turn implies that the $\Vbar$ state, which started at a point $J_2 =J_1/3$ at $h = h_{sat}$, extends over the whole range of $J_2$ near $h_{sat}/2$ (see Fig.~\ref{fig:PD}).

 \begin{table}[htb!]
\begin{center}
  \begin{tabular}{|c|c|c|c|c|c|}
    \hline
    $J_2/J_1$ & \quad$1/8$\quad\quad & \quad$1/4$\quad\quad & \quad$1/3$\quad\quad & \quad$1/2$\quad\quad & \quad$1$\quad\quad \\
    \hline
 $\delta_1  (1/S)$   &       0.46       &    0.15       &  0.11 & 0.11 & 0.28 \\\hline
 $\delta_2  (1/S)$ & 1.2 &   0.80       & 0.75 & 0.75 & 1.09\\
  \hline
  \end{tabular}
\end{center}
\caption{Results for the boundaries of UUUD state for different $J_2/J_1$ (see SM for details of calculations).
 The UUUD state is stable in the range $h_l<h<h_u$, where $h_l=h_{sat}/2-\delta_1$ and $h_u=h_{sat}/2+\delta_2$ . \label{tab:spectrumcorr}}
\end{table}

 At the lower boundary of the UUUD phase, two other spin-wave modes become unstable at the $\Gamma$ point. To determine the the spin order below $h_l$, we again perform  Landau Free energy analysis in terms of the corresponding two complex order parameters  $\Delta_1$ and $\Delta_2$. We present the details in the SM.
 The Free energy has the form~\cite{Nandkishore2012,Venderbos2016}:
\begin{align}
E_{\Delta}/N=&-\mu (|\Delta_1|^2+|\Delta_2|^2)+\frac{1}{2}\Gamma(|\Delta_1|^2+|\Delta_2|^2)^2\non\\&+\frac{1}{2}K|\Delta_1^2+\Delta_2^2|^2
\label{eq:QuarticFE}
\end{align}
Classically, $\Gamma=h_{sat}/4$, $K=0$. Then $|\Delta_1|^2+|\Delta_2|^2\equiv\mu/\Gamma$, i.e. different ordered states are degenerate.
 Quantum fluctuations lift the degeneracy, and the result depends on the sign of $K$. If $K>0$,  $\Delta_1=\pm\, i\Delta_2$.  It can be checked that this gives rise to a non-coplanar  umbrella state, in which the down-spin remains intact, and three up-spins split out and form a cone. Such a state breaks $U(1)\times\Zbb_4\times\Zbb_2$ symmetry. If $K<0$, the relative phase between $\Delta_1$ and $\Delta_2$ is either $0$ or $\pi$, and the order is coplanar (see Fig.~\ref{fig:plateauDown}).

We computed $K$ to accuracy $1/S$. The details of calculations are presented in SM, and here we quote the result: $K$ is the sum of logarithmical, $|\log (h_l-h)|/S$, $\log S/S$, and non-logarithmical, $\mc{O}(1/S)$ terms, much like Eq.~\ref{eq:dGamma}. The logarithmical term yields $K <0$, however the prefactor for the logarithm vanishes at $J_2 = J_1/3$, and at this value of $J_2$ non-logarithmical terms become relevant.  Near $J_2=J_1/3$, we have
\begin{align}
K=&-\frac{2\sqrt{3}J_1}{\pi}~\big(\frac{J_2}{J_1}-1/3\big)^2 \big(\frac{|\log (h_l-h)|}{S}+\beta_{\phi}\frac{\log S}{S}\big)\non\\&- \frac{\beta_K}{S}.
\label{eq:resK}
\end{align}
Where the $|\log (h_l-h)|/S$ term is a contribution from spin wave modes which go as $k^2$ at $h=h_l$, and $\log S/S$ term comes from another spin wave mode that softens at $h=h_u=h_l+\mc{O}(1/S)$. In distinction to the situation near $h_{sat}$, here we found that $K$ remains negative, even for  $J_2/J_1=1/3$. This implies that the state below $h_l$ is a co-planar state. An umbrella state is not ruled out, however, for smaller $S$ as we computed $\beta_K$ in Eq.~\ref{eq:resK} at $S \gg 1$.

To determine the structure of the coplanar state below $h_l$ more work is actually required because for $K <0$, the Free energy to order $\Delta^4$ is $E_{\Delta}/N=-\mu (|\Delta_1|^2+|\Delta_2|^2)+\frac{1}{2}(\Gamma-|K|)(|\Delta_1|^2+|\Delta_2|^2)^2$, i.e., the degeneracy is not fully lifted.  To select the order, one has to compute $\mc{O}(\Delta^6)$ terms in the Free energy. We found (see SM for detail) that sixth-order terms select the order in which of the three up-spins  two are tilting in  one direction and another in the opposite direction, while the down spin remains intact (see Fig.~\ref{fig:plateauDown}). This state breaks $U(1)\times\Zbb_4\times\Zbb_3$ symmetry. It can potentially transform gradually into the stripe state, which breaks $U(1)\times\Zbb_3$, if the down spin begins rotating at higher deviations from $h_l$ and match the spin from up-triad, which is separated from the other two. Or, the transition can be first order.  Either way, at small fields, the order becomes a stripe.  A more complex phase diagram at low fields is expected in the presence of a single-ion anisotropy~\cite{Seabra2011,Coldea2014}. 

\textbf{Conclusions\quad} In this work we analyzed the phase diagram of a Heisenberg antiferromagnet on a triangular lattice, with nearest and second nearest neighbor  interactions ($J_1-J_2$ model), in a magnetic field. We  focused on the case $1/8<J_1/ J_2 <1$, when semiclassical description involves four sublattice representation. We argued that the phase diagram is quite rich and contains several phases, besides the stripe state, which breaks $\Zbb_3$ orientational symmetry and has been detected at $h=0$ and near a saturation field. The most substantial result of our study is the identification of the UUUD phase, in which spins in the three sublattices are directed along the field, and spins in the fourth sublattice are directed opposite to the field. Such a state breaks a discrete $Z_4$ sublattice symmetry, but no orientational and continuous symmetry. As a result, all excitations in the UUUD phase are gapped, and the magnetization is fixed at exactly $1/2$ of the saturation value. We demonstrated that this phase is stable in a finite range of fields near $h_{sat}/2$ and is likely the true ground state of the model at all $J_2$ from the interval $1/8<J_1/ J_2 <1$.
We identified gapless planar states around the UUUD phase. The one at higher fields is the $\Vbar$ state. It breaks $U(1) \times \Zbb_4$ symmetry. The one at lower fields breaks  $U(1) \times \Zbb_4\times\Zbb_3$ symmetry.  A close competitor to this last state is a non-coplanar umbrella state.
 Such a state may potentially develop at a smaller $S$.  We call for magnetization measurements in quasi-2D triangular-lattice antiferromagnets with $J_2 > J_1/8$, best matetials with $S=1/2$, as they normally have no single-ion anisotropy, but also $S=1$ materials, like 2$H-$AgNiO$_2$~\cite{Coldea2014,Coldea2008}, to verify the existence of the plateau at a half of the saturation value of magnetization and quantum phases above this field.
 
\textbf{Acknowledgement\quad}
We acknowledge with thanks useful conversations with C. Batista, A. Coldea, R. Coldea, S-W Cheong, J. Kang, N. Perkins, and O. Starykh.  We are particularly thankful to R. Coldea for careful reading of the manuscript and useful comments. The work was supported by the NSF DMR-1523036.

\bibliography{bibJ1J2MY}
\onecolumngrid
\clearpage
\appendix
\section*{Supplementary Material}
In the Supplementary Material  we present technical details of calculations, which we reported in the Manuscript. We focus on the analysis of the half-magnetization plateau state (UUUD). Calculations right below the saturation field have been presented in our earlier work~\cite{Ye2017a}.

In the formulas below, $N$ is defined as the number of sites in a given sublattice, i.e. $N=\frac{N_{tot}}{n_{subl}}$. For example, for the four sublattice states, $n_{subl}=4$, and $N=\frac{1}{4}N_{tot}$.
\section{A: Low energy spectrum of the UUUD state}
\label{app:Hartree}
The excitation spectrum of the UUUD state can be straightforwardly obtained by using a four-sublattice representation for three spin-up and one spin-down sublattices and introducing four sets of Holstein-Primakoff (H-P) bosons $a,~b,~c$ for the spin-up sublattices and $d$ for the spin-down sublattice. The linear spin wave Hamiltonian reads:
\begin{align}
\mathcal{H}_{uuud}=&S\sum_{\kv}\{[(2\xi_{\kv}^{c}\ad_{\kv}b_{\kv}+2\xi_{\kv}^{a}\bd_{\kv}c_{\kv}+2\xi_{\kv}^{b}\cd_{\kv}a_{\kv})+h.c.]+[(2\xi_{\kv}^{a}\ad_{\kv}\dd_{-\kv}+2\xi_{\kv}^{b}\bd_{\kv}\dd_{-\kv}+2\xi_{\kv}^{c}\cd_{\kv}\dd_{-\kv})+h.c.]\non\\&+(-2h_0+h)(\ad_{\kv} a_{\kv}+\bd_{\kv} b_{\kv}+\cd_{\kv} c_{\kv})+(6h_0-h)\dd_{\kv}d_{\kv}\}
\label{eq:Huuud}
\end{align}
where $h_0\equiv(J_1+J_2)$. $\xi_{\kv}^{\alpha_i}~(\alpha_i=a,b,c)$ are the structure factors due to the exchange interactions between sublattices. They take the form of $\xi_{\kv}^{\alpha_i}=J_1\cos \kv\cdot\vect{\delta}_i+J_2\cos \kv\cdot\vect{l}_i$, where $\vect{\delta}_2=(1,0),\,\vect{\delta}_{1,3}=(\frac{1}{2},\mp\frac{\sqrt{3}}{2})$, $\vect{l}_2=(0,\sqrt{3}),\,\vect{l}_{1,3}=(\pm\frac{3}{2},\frac{\sqrt{3}}{2})$ are the nearest and next-nearest neighbor bonds respectively in unit of the lattice constant. To obtain the classical spin wave spectrum, i.e. to diagonalize the classical quadratic Hamiltonian of Eq.~\ref{eq:Huuud}, we write $\Hcal^{(2)}$ in the matrix form as: $\Hcal^{(2)}=\sum_{\kv}\Psi_{\kv}^{\dagger}H_{\kv}^{(2)}\Psi_{\kv}$.
\begin{align}
H_{\kv}^{(2)}=S
   \begin{pmatrix}
      -2h_0+h &  2\xi^{c}_{\kv} & 2\xi^{b}_{\kv} & 2\xi^{a}_{\kv} \\
      2\xi^{c}_{\kv} &  -2h_0+h & 2\xi^{a}_{\kv} & 2\xi^{b}_{\kv} \\
      2\xi^{b}_{\kv} & 2\xi^{a}_{\kv} &  -2h_0+h & 2\xi^{c}_{\kv} \\
      2\xi^{a}_{\kv} & 2\xi^{b}_{\kv}& 2\xi^{c}_{\kv} & 6h_0-h\\
  \end{pmatrix},
  \label{eq:quadraticH}
  \end{align}
where $\Psi_{\kv}=\{a_{\kv},b_{\kv},c_{\kv},\dd_{-\kv}\}^T$. To preserve the commutation relation of bosons, the canonical transformation $\Psi_{\kv}=T_{\kv}\Phi_{\kv}$ satisfies $g=TgT^{\dagger}$, where $g=\text{diag}~(1,1,1,-1)$\cite{Colpa1978}. As a result, $T_{\kv}$ that diagonalizes the Hamiltonian satisfies $T_{\kv}^{-1}g\,H^{(2)}T_{\kv}=g\,\Lambda_{\kv}$~\cite{Ye2017a}, where $\Lambda_{\kv}=\text{diag}\,(\omega_{\tilde a,\kv},\omega_{\tilde b,\kv},\omega_{\tilde c,\kv},\omega_{\tilde d,\kv})$. $\omega_{\kv}$ are the spin-wave spectrum, and $\Phi_{\kv}=\{\ta_{\kv},\tb_{\kv},\tc_{\kv},\tdd_{-\kv}\}^T$ are canonical eigenmodes.

Diagonalizing the linear spin wave Hamiltonian yields three gapless and one gapped spin wave branches. The latter one, with gap $\hsa/2$, describes the in phase precession of UUUD state around the magnetic field. All three \textit{quadratic} dispersing gapless modes soften at $\kv=0$, i.e. $\Gamma$ point of the four-sublattice Brillouin zone. In the following, we focus on the spectrum at $\Gamma$ point, and analyze the stability of the UUUD state and its proximate states. $H_{\kv}^{(2)}$ at the $\Gamma$ point can be diagonalized through a global rotation of basis,
\begin{equation}
\begin{pmatrix}
    a_{\kv} \\
    b_{\kv} \\
    c_{\kv}\\
    \dd_{-\kv}\\
\end{pmatrix}=
 \begin{pmatrix}
    \frac{1}{\sqrt{2}} &  \frac{-1}{\sqrt{6}} & \frac{1}{\sqrt{2}} & \frac{-1}{\sqrt{6}} \\
      \frac{-1}{\sqrt{2}} & \frac{-1}{\sqrt{6}} & \frac{1}{\sqrt{2}} & \frac{-1}{\sqrt{6}} \\
      0 & \frac{2}{\sqrt{6}} & \frac{1}{\sqrt{2}} &\frac{-1}{\sqrt{6}} \\
      0 & 0 & \frac{-1}{\sqrt{2}} & \frac{3}{\sqrt{6}} \\
  \end{pmatrix}
  \begin{pmatrix}
     e_{\kv} \\
     f_{\kv} \\
    \bar c_{\kv}\\
     \phi^{\dagger}_{-\kv}\\
      \end{pmatrix},
      \label{eq:RotUUUD}
\end{equation}
where $\{e_{\kv},f_{\kv},\phi^{\dagger}_{\kv}\}$ are the low energy modes. The spectrum at $\Gamma$ reads,
\begin{align}
\Hcal_0=&S\{(h-4h_0)\,(\ed_{\vect{0}}e_{\vect{0}}+\fd_{\vect{0}}f_{\vect{0}})+(4h_0-h)\, \phid_{\vect{0}}\phi_{\vect{0}}\},
\label{eq:HGamma}
\end{align}
which indicates that classically only at $h=4h_0=\hsa/2$, the spectrum of all spin-wave branches is non-negative and the UUUD state is among the generate ground state manifold. To see if quantum fluctuations stabilize the UUUD state in a range of field near $\hsa/2$, corrections to the spectrum at $1/S$ order should be calculated. As there is no 3-boson interactions for a collinear state in the isotropic Heisenberg model, the $1/S$ corrections to the spectrum only come from the 4-boson interactions through the 1/S Hartree-type self-energy. The 4-boson interaction term of the UUUD state reads:
\begin{align}
\Hcal^{(4)}=&\frac{4}{N}\sum_{\{\alpha,\beta,\gamma\},\kv_1-\kv_3}\{\frac{-1}{2}\big[\xi^{\gamma}_{1}(\ad_{\beta,1}\ad_{\alpha,2}a_{\alpha,3}a_{\alpha,1+2-3}+\ad_{\alpha,1}\ad_{\beta,2}a_{\beta,3}a_{\beta,1+2-3}+\dd_1\ad_{\gamma,2}\ad_{\gamma,3}a_{\gamma,1+2+3}\non\\&+\ad_{\gamma,1}\dd_2\dd_3 d_{1+2+3})+h.c.\big]+2\xi^{\gamma}_{1-2}\big(\ad_{\alpha,1}a_{\alpha,2}\ad_{\beta,3}a_{\beta,1+3-2}-\ad_{\gamma,1}a_{\gamma,2}\dd_{3}d_{1+3-2}\big) \}.
\label{eq:quarticH}
\end{align}
$N$ is the total number of sites. For brevity, we denote $\xi_{1}\equiv\xi_{\kv_1}$, $d_{1}\equiv d_{\kv_1}$, and so forth. Note that $\xi_{\kv}=\xi_{-\kv}$. The set of $\{a_{\alpha},a_{\beta},a_{\gamma}\}$ runs over all cyclic permutations of  $\{a,b,c\}$ as $\{a,b,c\}$, $\{c,a,b\}$, $\{b,c,a\}$.

The corrections to quadratic terms from $\mc{H}^{(4)}$ can be obtained by contracting two magnons, i.e. calculating magnon densities such as $\cond{\ad_{\alpha}a_{\beta}}_{0,\kv}$, $\cond{\xi\ad_{\alpha}a_{\beta}}_{0,\kv}$. The subscript of $\cond{...}_{0,\kv}$ labels the quantities averaged over, e.g. $0$ labels averaging over the quadratic Hamiltonian, $\kv$ averages over the crystal momentum in the sublattice Brillouin zone. The density averages can be obtained by the eigenvectors (columns of matrix $T$). For example,
\begin{align}
\cond{\ad_{\alpha}a_{\beta}}_{0,\kv}=\cond{(T_{\alpha'\alpha}\tad_{\alpha'}+T_{4\alpha}\td)(T_{\beta'\beta}\ta_{\beta'}+T_{4\beta}\tdd)}_{0,\kv}=\frac{1}{N}\sum_{\kv\in B.Z.}T_{4\alpha}T_{4\beta}\equiv\cond{T_{4\alpha}T_{4\beta}}_{\kv}
\end{align}
$\cond{\ad_{\alpha}\dd}_{0,\kv}=\cond{T_{4\alpha}T_{44}}_{\kv}$, $\cond{a_{\alpha}d}_{0,\kv}=\cond{T_{1\alpha}T_{14}+T_{2\alpha}T_{24}+T_{3\alpha}T_{34}}_{\kv}=\cond{T_{4\alpha}T_{4\beta}}_{\kv}$, $\cond{\dd d}_{0,\kv}=\cond{T_{14}^2+T_{24}^2+T_{34}^2}_{\kv}=-1+\cond{T_{44}^2}_{\kv}$, and so forth. $g=TgT^{\dagger}$ has been applied to simplify the above expressions. As the matrix $T_{\kv}$ is a regular function of $\kv$, the self-energy should also be regular near $\Gamma$. Thus it is enough to calculate the Hartree terms at $\Gamma$, where classical spin-wave modes soften. Expressing them in terms of the canonical eigenmodes $\{e_{\vect{0}},f_{\vect{0}},\phi^{\dagger}_{\vect{0}}\}$, we have
\begin{align}
\delta \mathcal{H}_0=S \{\delta_1\,(\ed_{\vect{0}}e_{\vect{0}}+\fd_{\vect{0}}f_{\vect{0}})+\delta_2\, \phid_{\vect{0}}\phi_{\vect{0}}\}.
\label{eq:deltaHGamma}
\end{align}
where $\delta_1,\,\delta_2$ are linear combinations of the averages. They sets the boundary of the UUUD state by requiring all the three modes are gapped at $\Gamma$, i.e. $h-4h_0+\delta_1>0$, $4h_0-h+\delta_2>0$. From $h>4h_0-\delta_1$, we define the lower critical field as $h_l=4h_0-\delta_1$, below which $\{e,\,f\}$ modes soften. From $h<4h_0+\delta_2$, we obtain the upper critical field above which $\phi$ mode softens. The numerical values of $\delta_1,\,\delta_2$ at different $J_2/J_1$ are listed in Table~I. As $\delta_1,\,\delta_2$ are positive for all $J_2/J_1$ in the range $1/8<J_2/J_1<1$, the UUUD state is stable in this full range of $J_2/J_1$. 

The pattern of transverse magnetic order above $h_u$ and below $h_l$ can be identified straightforwardly from Eq.~\ref{eq:RotUUUD}. Above $h_u$, the $\phi$ mode develops condensate. From Eq.~\ref{eq:RotUUUD}, 
\begin{equation}
\langle a_0\rangle=\frac{-1}{\sqrt{6}}\langle \phid_0\rangle,~\langle b_0\rangle=\frac{-1}{\sqrt{6}}\langle \phid_0\rangle,~\langle c_0\rangle=\frac{-1}{\sqrt{6}}\langle \phid_0\rangle,~\langle \dd_0\rangle=\frac{3}{\sqrt{6}}\langle \phid_0\rangle
\end{equation}
Due to the rotation symmetry along the field, $\cond{\phi_0}=|\cond{\phi_0}|e^{i\varphi}$, $\varphi\in(0,2\pi)$. We relate the spin order with the magnon condensate through the H-P transformation, i.e. $\cond{S_{\alpha,x}}=\sqrt{2S}\frac{\cond{a_\alpha}+\cond{\ad_\alpha}}{2}=-\sqrt{\rho}\cos \varphi$, $\cond{S_{\alpha,y}}=\sqrt{2S}\frac{\cond{a_\alpha}-\cond{\ad_\alpha}}{2i}=-\sqrt{\rho}\sin \varphi$ for $\{a,b,c\}$ and $\cond{S_{d,x}}=\sqrt{2S}\frac{\cond{d_\alpha}+\cond{\dd_\alpha}}{2}=3\sqrt{\rho}\cos \varphi$, $\cond{S_{d,y}}=-\sqrt{2S}\frac{\cond{d_\alpha}-\cond{\dd_\alpha}}{2i}=3\sqrt{\rho}\sin \varphi$ for $d$ sublattice, where $\rho\equiv\frac{S}{3}|\cond{\phi_0}|^2$. Thus the transverse spin order has three up-spins point to one direction and the down-spin point to the opposite direction, which matches the spin order of the $\Vbar$ state near $\hsa/2$. Fixing the phase of the condensate $\varphi$ specifies the plane of the $\Vbar$ state and breaks $U(1)$ rotation symmetry around the field.

\begin{figure}[htb!]
  \centering
  \subfigure[]{\includegraphics[width=0.3\linewidth]{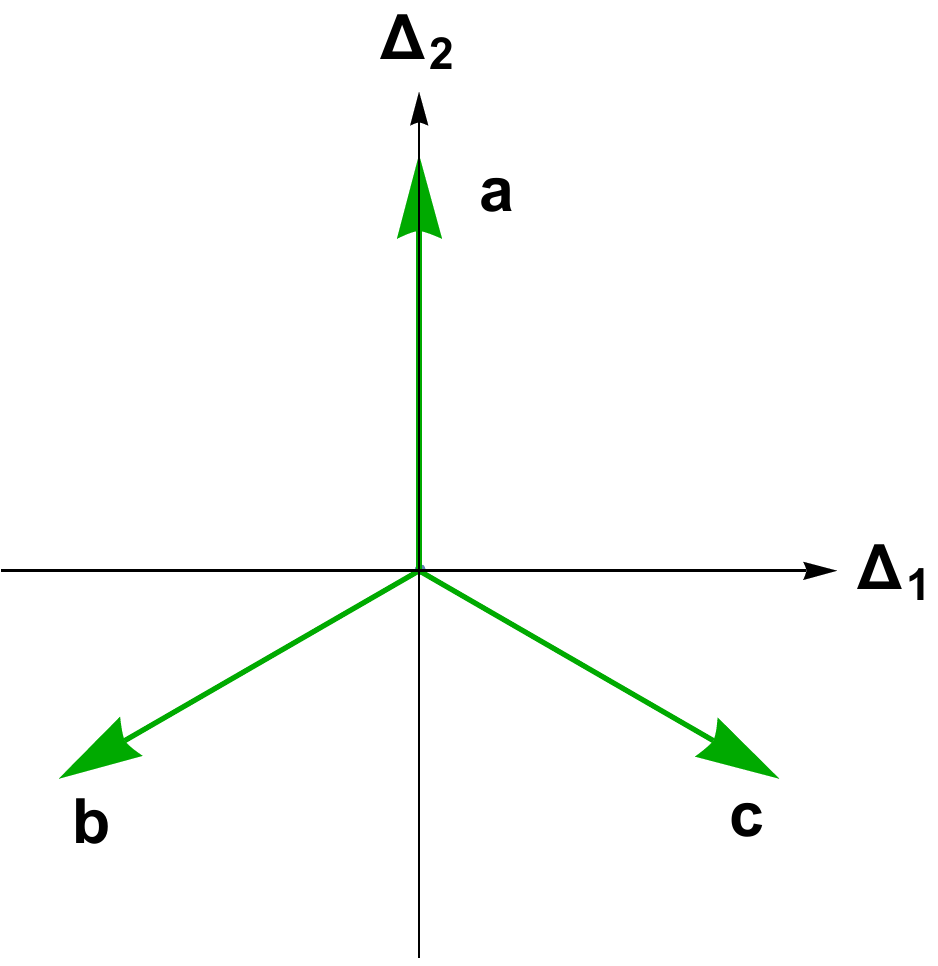}\label{figChiralOD1}}\quad
  \subfigure[]{\includegraphics[width=0.3\linewidth]{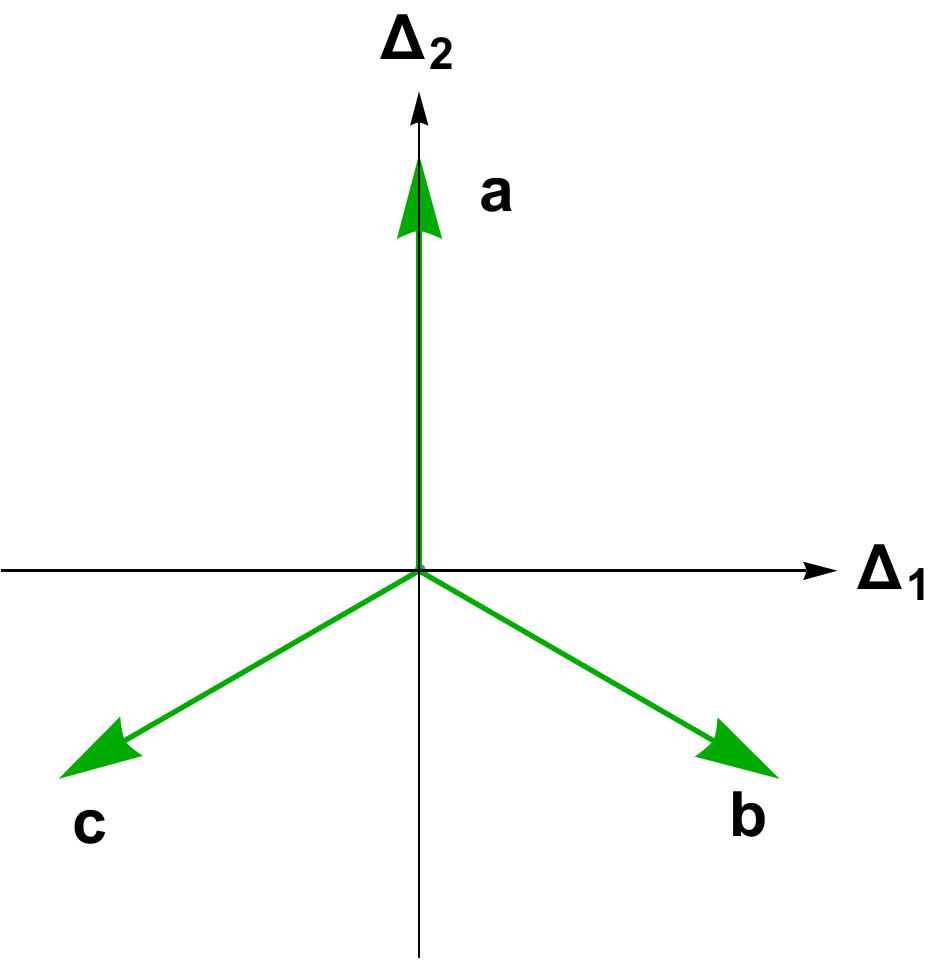}\label{figChiralOD2}}\quad
  \subfigure[]{\includegraphics[width=0.3\linewidth]{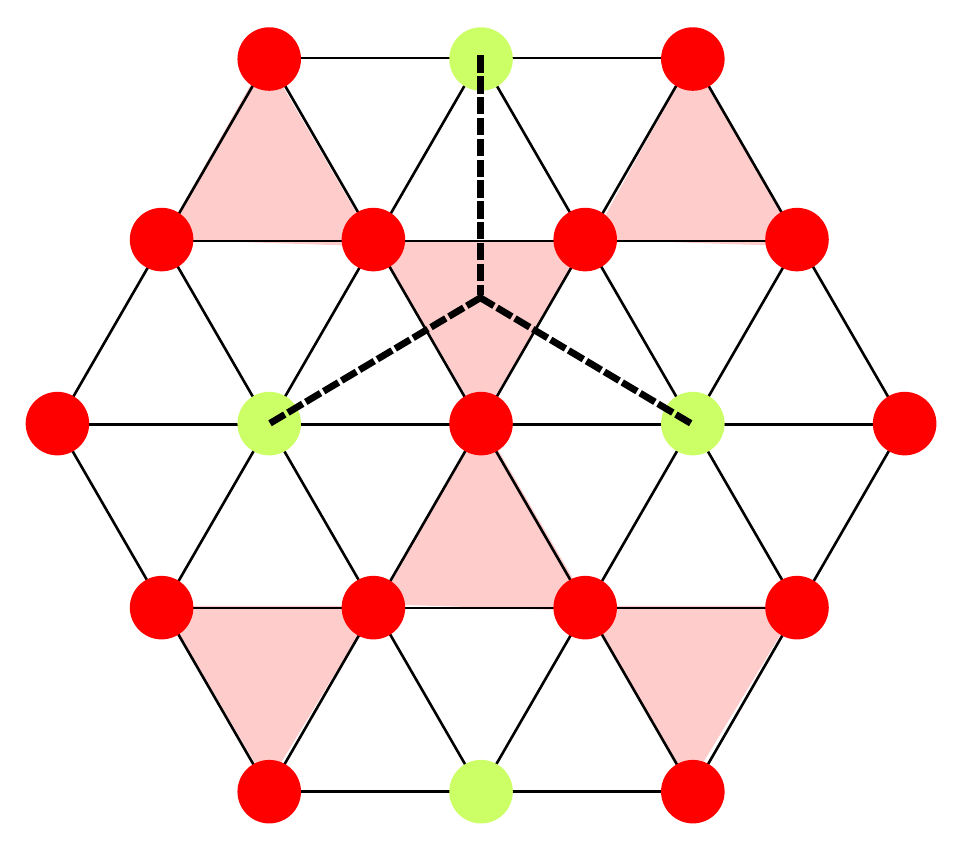}}
  \caption{(a), (b) Transverse order of the umbrella state, $|\Delta_1|\pm i|\Delta_2|$. As the down-spin align with the field, there is no transverse order associated with it. (c) Illustration of lattice symmetry of the UUUD state. Darker (red) sites for up-spin sublattices, Ligher (green) sites for the down-spin sublattice. The rotation center of $C_{3v}$ space group is at the center of a given shaded (red) triangle. The reflection is around the dashed lines. 
  \label{fig:chiralOD}}
  \end{figure}

The pattern of the transverse magnetic order below $h_l$ is more complex due to the degeneracy of the zero modes $\{e_{\vect{0}},\,f_{\vect{0}}\}$ at $h_l$. And the order can be non-coplanar if the relative phase between condensates $\Delta_1\equiv\cond{e_0}$ and $\Delta_2\equiv\cond{f_0}$ is non-zero mod $\pi$. In particular, if $\Delta_2=\pm\, i\Delta_1$, the transverse orders associated with $\Delta_1$ and $\Delta_2$ takes a relative angle $\pm \pi/2$, thus the transverse order of $|\Delta_1|\pm i|\Delta_2|$ form a equilateral triangle as shown in Fig.~\ref{figChiralOD1} and (b), where $\pm$ interchanges $b,\,c$ sublattice labels. And the total magnetic order (transverse plus longitudinal) has non-zero chirality, i.e. an umbrella state.
  
\section{B: Symmetry constraint on the Landau free energy}
\label{app:LandauFE}
We show details of obtaining the form of the Landau free energy in powers of magnon condensates $\Delta$ below $h_l$, the lower critical field at which the UUUD state becomes unstable. As explained at the end of Sec. A, the order below $h_l$ should be described by a two-component order parameter, two degenerate The Landau free energy of $\Delta$ should respect a $C_{3v}$ symmetry, i.e. threefold rotation around the center of a triangle formed by three nearest neighbor up-spins ($C_3$) and three reflections in the symmetry lines of the equilateral triangle ($\sigma_v$). Upon lowering the field, the only classical ground state configuration that respect the $C_{3v}$ symmetry is umbrella state, which breaks $\Zbb_2$ chiral symmetry. The transverse magnetic order of spin $\alpha$ ($\alpha=a,b,c$) can be expressed as $\Delta_+ e^{i\vect{Q}\cdot \vect{R}_\alpha}$ or $\Delta_- e^{-i\vect{Q}\cdot \vect{R}_\alpha}$, where $\Delta_{\pm}$ denotes the spiral order of different chirality, $\vect{R}_{a}=(0,0)$, $\vect{R}_b=(-\frac{1}{2},\frac{\sqrt{3}}{2})$, $\vect{R}_c=(-\frac{1}{2},-\frac{\sqrt{3}}{2})$. $\{e^{i\vect{Q}\cdot \vect{R}_\alpha}, e^{-i\vect{Q}\cdot \vect{R}_\alpha}\}$ forms the basis of the $\Gamma_3$ irreducible representation of $C_{3v}$ symmetry group. Thus a generic order parameter in $\Gamma_3$ representation is:
\begin{align}
\Delta_{\alpha}=\frac{1}{2\sqrt{3}}\big(\Delta_+ e^{i\vect{Q}\cdot \vect{R}_\alpha}+\Delta_- e^{-i\vect{Q}\cdot \vect{R}_\alpha}\big)
=\Delta_1\vect{v}_{1\alpha}+\Delta_2\vect{v}_{2\alpha}
\end{align}
where $\vect{v}_1=\{0,\frac{-1}{\sqrt{2}},\frac{1}{\sqrt{2}}\}$, $\vect{v}_2=\{\frac{2}{\sqrt{6}},\frac{-1}{\sqrt{6}},\frac{-1}{\sqrt{6}}\}$. $\{\vect{v}_{1},\vect{v}_{2}\}$ forms another basis of the $\Gamma_3$ representation, and it is defined in such a way that $\Delta_1,\Delta_2$ is the same as defined in previously $\Delta_1\equiv\cond{e_0}$ and $\Delta_2\equiv\cond{f_0}$ up to exchanging the sublattice labels $a,\, c$. $\Delta_\pm$ and $\Delta_{1,2}$ are related as $\Delta_1=\frac{-i}{\sqrt{2}}\,(\Delta_+-\Delta_-)$, $\Delta_2=\frac{1}{\sqrt{2}}(\Delta_++\Delta_-)$. The Landau free energy at order $\Delta_\pm^{2n}$ can be obtained by finding all channels that contribute to $(\Gamma_3\otimes\Gamma_3)^n\rightarrow \mathbb{I}$.

At quadratic order ($n=1$), the only term is $|\Delta_+|^2+|\Delta_-|^2\equiv|\Delta_1|^2+|\Delta_2|^2$.

At quartic order, as $(\Gamma_3\otimes\Gamma_3)^2=(\Gamma_1\oplus\Gamma_2\oplus\Gamma_3)^2$, there are three channels contributing to identity. From $\Gamma_1\otimes\Gamma_1$, one can get $(|\Delta_+|^2+|\Delta_-|^2)^2\equiv (|\Delta_1|^2+|\Delta_2|^2)^2$; from $\Gamma_2\otimes\Gamma_2$, one can get $(|\Delta_+|^2-|\Delta_-|^2)^2\equiv |\Delta_1^*\Delta_2-\Delta_1\Delta_2^*|^2$; from $\Gamma_3\otimes\Gamma_3$, one get $(|\Delta_+^*\Delta_-|^2+|\Delta_-^*\Delta_+|^2)\equiv |\Delta_1^2+\Delta_2^2|^2$. Combining all three contributions, the quartic term takes the form of Eq.~4, which can break the $\Zbb_2$ ($\sigma_v$) chiral symmetry and select $\Delta_+$ or $\Delta_-$, but cannot break $\Zbb_3$ ($C_3$) symmetry. 

We show that $\Zbb_3$ symmetry can be spontaneously broken by the six-order term. From the channel $(\Gamma_3\otimes\Gamma_3)^3\rightarrow(\Gamma_3)^3\rightarrow\Gamma_3\otimes\Gamma_3\rightarrow \mathbb{I}$, one can obtain terms of $\Delta_+^3\Delta_-^{*3}+\Delta_+^{*3}\Delta_-^{3}$ and $\Delta_+^6+\Delta_-^6$. In terms of $\Delta_1$ and $\Delta_2$, when their relative phase is $0$ or $\pi$, the above two terms take the same form of $\Delta_1^6 - 15 \Delta_1^4 \Delta_2^2 + 15 \Delta_1^2 \Delta_2^4 - \Delta_2^6$, which select either $\Delta_1$ or $\Delta_2$ and thus spontaneously breaks the $\Zbb_3$ symmetry.

It turns out that for a system with certain trigonal or hexagonal symmetry, i.e. $C_{6v}$, $D_{6h}$, etc., the Landau free energy of a two-component order parameter takes the same form~\cite{Nandkishore2012,Venderbos2016}. Technically, it's because the multiplication table~\cite{Koster1963} for the two-dimensional irreducible representation of these symmetry groups are the same.
\section{C: Calculation of quartic coefficient $K$}
\label{app:LandauFreeEK}

Below $h_l$, both $e$ and $f$ magnon potentially softens. We  split the magnon operators $e_{\kv}$ and $f_{\kv}$ into condensate and non-condensate as
\begin{align}
e_{\kv}=\sqrt{N}\Delta_1\delta_{\kv,0}+\tilde e_{\kv},\quad
f_{\kv}=\sqrt{N}\Delta_2\delta_{\kv,0}+\tilde f_{\kv}
\label{eq:efDelta}
\end{align}
or in terms of $\{a_{\kv},b_{\kv},c_{\kv} \}$ as
\begin{align}
a_{\kv}=&\frac{1}{\sqrt{2}}\sqrt{N}\Delta_1\delta_{\kv,0}-\frac{1}{\sqrt{6}}\sqrt{N}\Delta_2\delta_{\kv,0}+\tilde a_{\kv}\qquad&
b_{\kv}=&-\frac{1}{\sqrt{2}}\sqrt{N}\Delta_1\delta_{\kv,0}-\frac{1}{\sqrt{6}}\sqrt{N}\Delta_2\delta_{\kv,0}+\tilde b_{\kv}\non\\
c_{\kv}=&\frac{2}{\sqrt{6}}\sqrt{N}\Delta_2\delta_{\kv,0}+\tilde c_{\kv}\qquad&
d_{\kv}=&\tilde d_{\kv}
\label{eq:abcDelta}
\end{align}
From (\ref{app:LandauFE}), the ground state energy density in powers of magnon condensates up to quartic order is:
\begin{align}
E_{\Delta}/N=&-\mu (|\Delta_1|^2+|\Delta_2|^2)+\frac{1}{2}\Gamma(|\Delta_1|^2+|\Delta_2|^2)^2+\frac{1}{2}K|\Delta_1^2+\Delta_2^2|^2
\label{eq:QuarticFEapp}
\end{align}
As explained in the manuscript, the condensates manifold that minimized the energy depends on the sign of $K$. In the following, we show how the sign of $K$ is obtained up to order $1/S$.

For convenience, Eq.~\ref{eq:quarticH} can be expressed as
\begin{align}
\mc{H}^{(4)}=\frac{1}{N}\sum_{i,j,l,k}V_{i,j,l,k}(\kv_1+\qv,\kv_2-\qv,\kv_1,\kv_2)\psid_{i,\kv_1+\qv}\psid_{j,\kv_2-\qv}\psi_{l,\kv_1}\psi_{k,\kv_2}.
\end{align}
where $\{\psi_{i,\kv}\}=\{a_{\kv},b_{\kv},c_{\kv},\dd_{-\kv}\}$ or $\{e_{\kv},f_{\kv},\tilde c_{\kv},\phid_{-\kv}\}$, depending on the basis of canonical modes in the context. $V_{i,j,l,k}$ is chosen such that double counting has been avoided. 

The classical value for $K$ is obtained by expressing all magnon operators by the condensates $\Delta_1,\,\Delta_2$ following Eq.~\ref{eq:efDelta} or Eq.~\ref{eq:abcDelta}. We find $K^{(0)}\equiv\frac{1}{2}V_{ffee}(\vect{0},\vect{0},\vect{0},\vect{0})=0$.

We now calculate $K^{(1)}$ from quantum corrections at order $1/S$. Only the sign of $K$ is significant, which generally doesn't change across the critical field $h_l$ below which the quadratic term becomes negative. We calculate the 4-boson interaction at order $1/S$ at $h_l$. There are two sources of quantum corrections, one from the normal ordering of the Holstein-Primakoff bosons, another from quantum fluctuations at second order perturbation.

\subsection{Corrections from normal ordering}
The Holstein-Primakoff (H-P) transformation
\begin{align}
S_{\vect{r}}^z(\vect{z})  =S-a_{\vect{r}}^{\dagger}a_{\vect{r}},
S_{\vect{r}}^+(\vect{z})  =\sqrt{2S}\sqrt{1-\frac{a_{\vect{r}}^{\dagger}a_{\vect{r}}}{2S}}a_{\vect{r}},
S_{\vect{r}}^-(\vect{z})  =\sqrt{2S}a_{\vect{r}}^{\dagger}\sqrt{1-\frac{a_{\vect{r}}^{\dagger}a_{\vect{r}}}{2S}}
\label{eq:HP}
\end{align}
contains $\sqrt{1-a_{\vect{r}}^{\dagger}a_{\vect{r}}/2S}$. To express the Hamiltonian in terms of the H-P bosons, we expand $\sqrt{1-a_{\vect{r}}^{\dagger}a_{\vect{r}}/2S}$ in powers of the bosons. Due to the normal ordering of the bosons in the expansion, e.g. $(\ad_ra_r)^2=\ad_r\ad_r a_r a_r+\ad_r a_r$, $S_{\vect{r}}^+$ can be written as:
\begin{align}
S_{\vect{r}}^+  =\sqrt{2S}(1-\frac{1}{4S}(1+\frac{1}{8S}+\frac{1}{32S^2}+...)a_{\vect{r}}^{\dagger}a_{\vect{r}})a_{\vect{r}}+\mathcal{O}(a^5)
\label{eq:HPexp}
\end{align}
Eq.~\ref{eq:quarticH} is obtained keeping the leading term in $1/S$, i.e. $S_{\vect{r}}^+ \approx\sqrt{2S}(1-\frac{1}{4S}a_{\vect{r}}^{\dagger}a_{\vect{r}})a_{\vect{r}}+\mathcal{O}(a^5)$, etc..
To obtain the 4-boson interaction to the second order in $1/S$, we keep to order $1/S^2$ in Eq.~\ref{eq:HPexp}, i.e. $S_{\vect{r}}^+ \approx\sqrt{2S}(1-\frac{1}{4S}(1+\frac{1}{8S})a_{\vect{r}}^{\dagger}a_{\vect{r}})a_{\vect{r}}+\mathcal{O}(a^5)$.

The normal ordering contribution to the quartic term reads:
\begin{align}
\delta\Hcal^{(4)}=&\frac{1}{8S}\frac{1}{N}\sum_{\alpha,\kv_1-\kv_3}\{\frac{-1}{2}\big[\xi^{\gamma}_{1}(\ad_{\beta,1}\ad_{\alpha,2}a_{\alpha,3}a_{\alpha,1+2-3}+\ad_{\alpha,1}\ad_{\beta,2}a_{\beta,3}a_{\beta,1+2-3}\non\\&+\dd_1\ad_{\gamma,2}\ad_{\gamma,3}a_{\gamma,1+2+3}+\ad_{\gamma,1}\dd_2\dd_3 d_{1+2+3})+h.c.\big]\}
\label{eq:quarticHnod}
\end{align}
Plug Eq.~\ref{eq:abcDelta} into Eq.~\ref{eq:quarticHnod}, the normal ordering contribution to $K$ is $K^{(1)}_a=\frac{J_1+J_2}{24S}$.

\subsection{Corrections from quantum fluctuations}
\begin{figure}[htb!]
  \centering
  \includegraphics[width=0.6\linewidth]{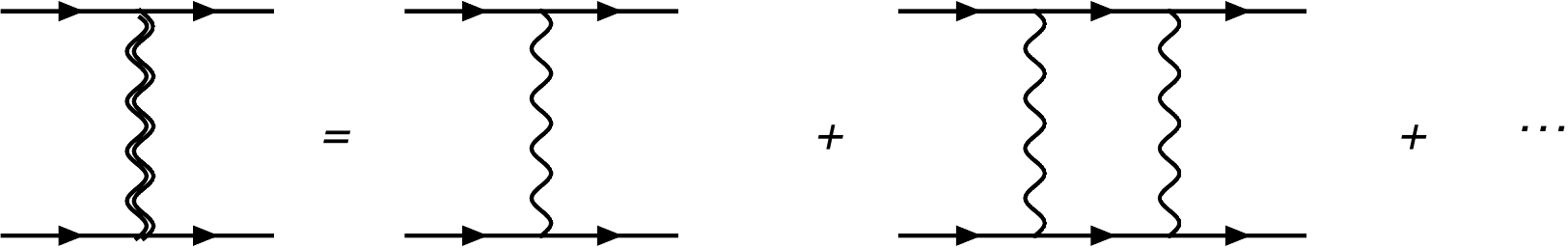}
  \caption{Diagrammatic representation of the perturbative corrections to the four-boson interaction.
  \label{fig:4ptvertex}}
  \end{figure}

Corrections to the 4-boson interaction at order $1/S$ can be expressed diagrammatically as shown in Fig.~\ref{fig:4ptvertex}. Following the Feynman rules that associate each 4-boson vertex with $-iV$ and each boson propagator with $iG_k=\frac{i}{\omega-\omega_{\kv}}$, the 4-boson interaction up to $1/S$ is
\begin{align}
-iV(\kv_1+\qv,\kv_2-\qv,\kv_1,\kv_2)&=-iV^{(0)}(\kv_1+\qv,\kv_2-\qv,\kv_1,\kv_2)+(-i)^2i^2\int_k\big(G_{k_1+k}G_{k_2-k}\non\\&V^{(0)}(\kv_1+\kv,\kv_2-\kv,\kv_1,\kv_2)V^{(0)}(\kv_1+\qv,\kv_2-\qv,\kv_1+\kv,\kv_2-\kv) \big)\non
\end{align}
As each ladder contributes a $1/S$ factor due to the magnon dispersion $\omega\sim S$, $1/S$ corrections to the 4-boson interaction and magnon dispersion is not relevant. The subscript of $G_k$, $\int_k$ is short for $k=(\omega,\kv)$. The indices for multiple magnon branches and the corresponding symmetry factor are suppressed for brevity.   $V^{(1)}(\kv_1+\qv,\kv_2-\qv,\kv_1,\kv_2)$ can be expressed as
\begin{align}
V^{(1)}=&i\int_kG_{k_1+k}G_{k_2-k}V^{(0)}(\kv_1+\kv,\kv_2-\kv,\kv_1,\kv_2)V^{(0)}(\kv_1+\qv,\kv_2-\qv,\kv_1+\kv,\kv_2-\kv)\non\\
=&-\int_{\kv}\frac{V^{(0)}(\kv_1+\kv,\kv_2-\kv,\kv_1,\kv_2)V^{(0)}(\kv_1+\qv,\kv_2-\qv,\kv_1+\kv,\kv_2-\kv)}{\omega_{\kv_1+\kv}+\omega_{\kv_2-\kv}}
\label{eq:secondpertV}
\end{align}
To get the expression in the second line, we integrate the frequency using
\begin{align}
\int_{\omega}\frac{\mathrm{d}\omega}{2\pi}\frac{1}{(\omega-\omega_{\kv_1})(-\omega-\omega_{\kv_2})}=\frac{-i}{\omega_{\kv_1}+\omega_{\kv_2}}.
\end{align}
From the expression of Eq.~\ref{eq:secondpertV}, one can see that when $\omega_{\kv_1+\kv}+\omega_{\kv_2-\kv}\sim S k^2$, the integral is logarithmical divergent if $V^{(0)}\sim\mc{O}(1)$. In the following, we first calculate the logarithmical part, which can be done analytically, and then calculate the regular $1/S$ part if necessary.
  
\begin{figure}[htb!]
  \centering
  \subfigure[]{\includegraphics[width=0.45\linewidth]{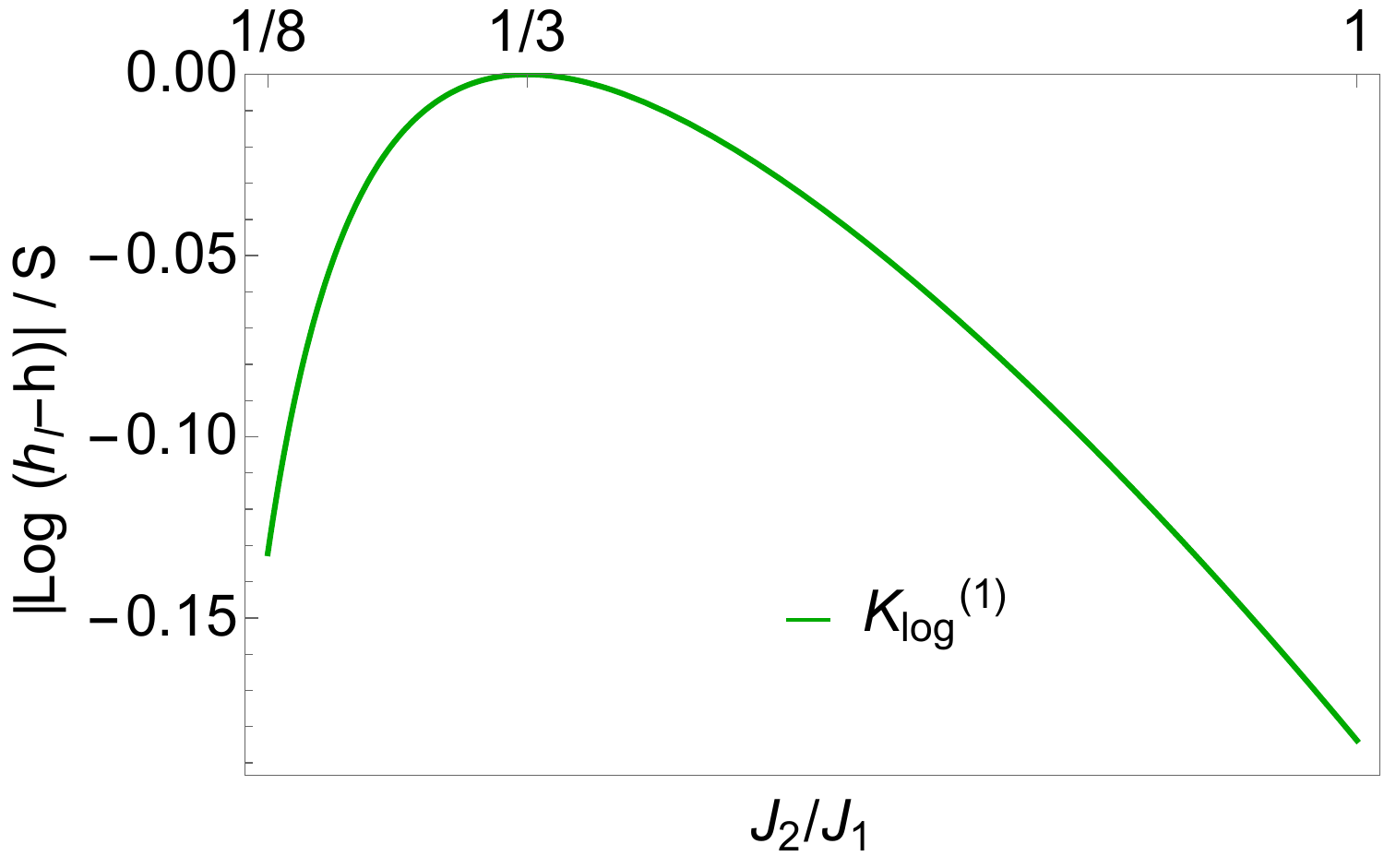}\label{fig:UUUDlog}}\quad
  \subfigure[]{\includegraphics[width=0.45\linewidth]{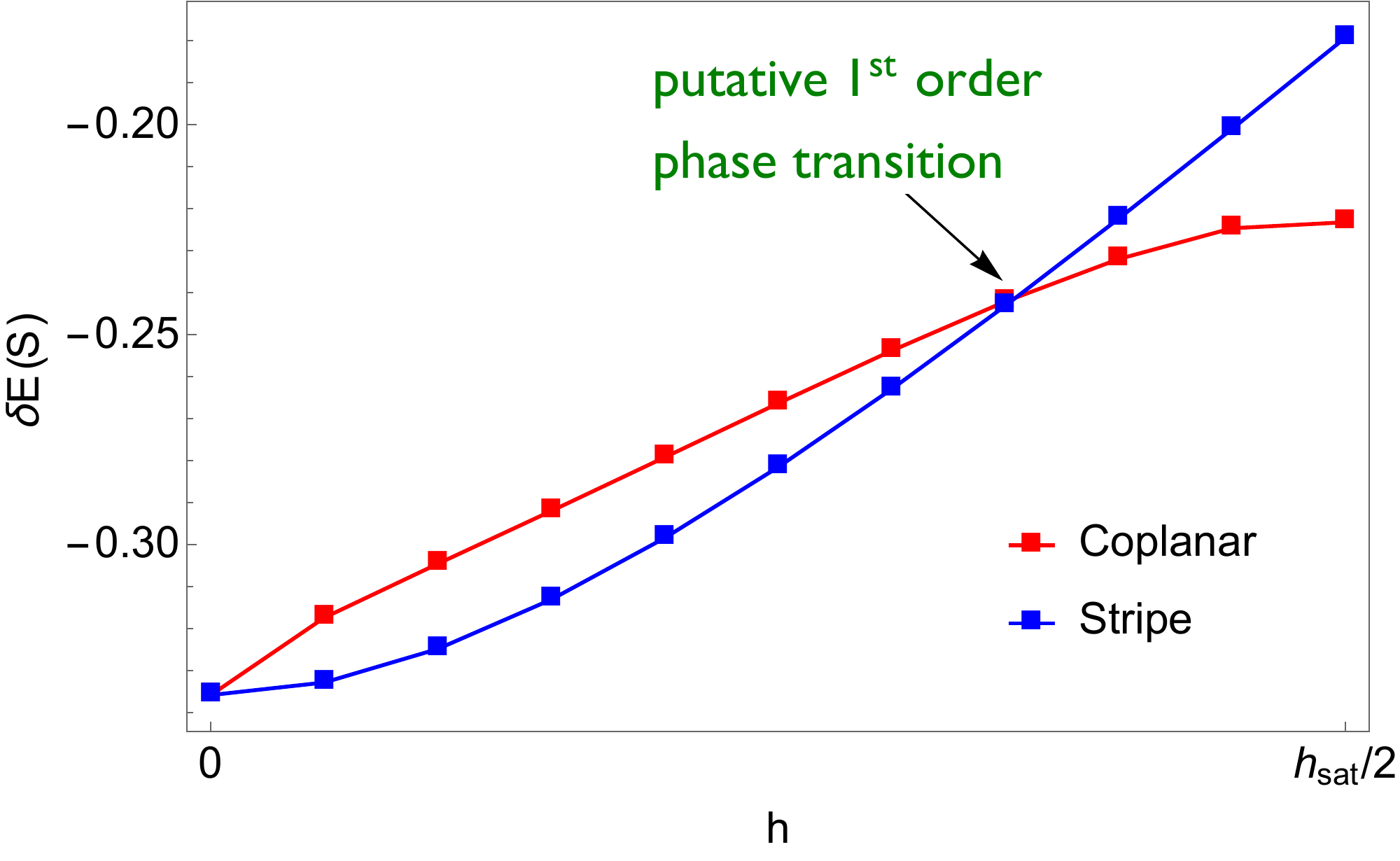}\label{fig:Eplot}}
  \caption{(a) $K^{(1)}$ at primary logarithmical accuracy, $|\log (h_l-h)|/S$. (b) $1/S$ correction to the ground state energy (zero point fluctuations) of the coplanar and stripe state for $0<h<\hsa/2$. The kink in the ground state energy indicates a first order phase transition between the coplanar state and canted stripe (The figure takes of value of $J_2=J_1/3$ as an example).}
\end{figure}

$K^{(1)}$ at logarithmical accuracy come from quadratic dispersive soft modes in the low energy $\{e,~f,~\phi\}$ sector (see Eq.~\ref{eq:RotUUUD}). First, we diagonalize the low energy spectrum at $h=h_l$ as follows. In terms of $\{e,f,\phid\}$, the low energy quadratic Hamiltonian can be written in two parts as $\Hcal_{uuud}=\Hcal_0+(J_2/J_1-1/3)\Hcal'$. $\Hcal_0$ is diagonal in $\{e,f,\phid\}$, while $\Hcal'$ mixes $\{e,f,\phid\}$ modes.
\begin{align}
\Hcal_0=S\sum_{\kv}\{k^2(\ed_{\kv}e_{\kv}+\fd_{\kv}f_{\kv})+(2k^2+\delta h)\, \phid_{\kv}\phi_{\kv}\},\non
\end{align}
where $\delta h=\delta_1+\delta_2=\mc{O}(1/S)$ is the width of the UUUD state. In the matrix form, $\Hcal'=S\sum_{\kv}\Psi^{\dagger}_{\kv}H'\Psi_{\kv}$, $\Psi_{\kv}=\{e_{\kv},f_{\kv},\phi^{\dagger}_{-\kv}\}^T$,
  \begin{equation}
  H'=k^2
   \begin{pmatrix}
    \frac{3}{2}(1+\cos\varphi_{\kv}) &  \frac{3}{2}\sin\phik & -3\sin\phik \\
      \frac{3}{2}\sin\phik & \frac{3}{2}(1-\cos\phik) & 3\sin\phik \\
      -3\sin\phik & 3\sin\phik & 3 \\
  \end{pmatrix}\non,
\end{equation}
where $\phik=2\theta_{\kv}+\pi/3$, and $\theta_{\kv}$ is defined as the angle between $\kv$ and the positive x-axis. When $k^2\ll 1/S$, the $\phi$ mode decouples from $e,~f$ modes that are gapless at $h_l$. Below $h_l$, the condensates develop, the quadratic dispersing gapless modes either acquire a gap that scales as $h_l-h$ or become linear dispersing Goldstone mode, both of which introduce a natural cutoff to the logarithm -- $|\log (h_l-h)|/S$ as a primary contribution at logarithmic accuracy. While $\phi$ mode starts to contribute to $K^{(1)}$ significantly when $k^2\gtrsim 1/S$ in the integral, at order $\log S/S<<|\log (h_l-h)|/S$. To obtain $K$ at order $|\log (h_l-h)|/S$, it is enough to use the 4-boson interaction in terms of $e,~f$ near $\kv=0$.
\begin{align}
\Hcal_l=h_0\frac{1}{N}\sum_{|\kv|<\Lambda}(\ed_{\kv}\ed_{-\kv} e_{\qv}e_{-\qv}+\fd_{\kv}\fd_{-\kv} f_{\qv}f_{-\qv}+2\ed_{\kv}\fd_{-\kv} e_{\qv}f_{-\qv})
\end{align}
where $h_0\equiv J_1+J_2$. Though there is no $\ed\ed f f+h.c.$ term at classical level, as $\{e,f\}$ are not the eigenmodes at $\kv\neq0$ when $J_2/J_1\neq1/3$, $\ed\ed f f$ at order $|\log (h_l-h)|/S$ is expected to be non-zero. First, condense two incoming or outgoing magnon modes and the non-condensed ones will be served as internal propagators, $\Hcal_l$ yields
\begin{equation}
\Hcal_{i,\kv}=h_0\frac{1}{N}\sum_{\kv}(\ed_{\kv}\ed_{-\kv} \Delta_1^2+\fd_{\kv}\fd_{-\kv} \Delta_2^2+2\ed_{\kv}\fd_{-\kv} \Delta_1\Delta_2)+h.c.
\label{eq:QuarticHi}
\end{equation}
The $\kv$-dependence of the interaction, not relevant at logarithmic accuracy, has been suppressed. Following Eq.~\ref{eq:secondpertV}, the leading order correction to the energy in proportion to $\Delta_1^2\bar{\Delta}_2^2+h.c.$ is:
\begin{align}
\Delta E_K=-h_0^2\Delta_1^2\bar{\Delta}_2^2\frac{1}{N}\sum_{\kv,\qv}\langle \ed_{\kv}\ed_{-\kv}f_{\qv}f_{-\qv}\rangle_0+h.c.
\label{eq:Hquartic}
\end{align}
where $\cond{\hat{O}}_0$ is the average of operator $\hat{O}$ over the quadratic Hamiltonian, which is equivalent to the boson scattering problem presented diagrammatically in Fig.~\ref{fig:4ptvertex}. Though different from the fully polarized state~\cite{Chubukov2014,Ye2017a}, the quadratic Hamiltonian of the UUUD state gets corrections from quantum fluctuations, the calculation can be self-consistently carried out in powers of $1/S$. For our purpose, it is enough to calculate $K$ up to order $1/S$. Comparing Eq.~\ref{eq:QuarticFEapp} and Eq.~\ref{eq:Hquartic}, $K=-2h_0^2\frac{4}{N}\sum_{\kv,\qv}\langle \ed_{\kv}\ed_{-\kv}f_{\qv}f_{-\qv}\rangle_0$. To calculate $\langle \ed_{\kv}\ed_{-\kv}f_{\qv}f_{-\qv}\rangle_0$, we obtain the canonical eigenmodes $u,~v$ at low energy through a rotation of basis:
\begin{align}
e_{\kv}&=\cos\tthk u_{\kv}-\sin\tthk v_{\kv}\non\\
f_{\kv}&=\sin\tthk u_{\kv}+\cos\tthk v_{\kv}
\label{eq:efRot}
\end{align}
where $\tthk=\thk-\pi/3$. The low energy spectrum of $u,~v$ modes are $\omega_{u,\kv}=S J_1 k^2$, $\omega_{v,\kv}=SJ_1(1+3\alpha)k^2$, where $\alpha\equiv J_2/J_1-1/3$. $\langle \ed_{\kv}\ed_{-\kv}f_{\qv}f_{-\qv}\rangle_0$ in terms of canonical eigenmodes $u$ and $v$ is:
\begin{align}
\langle \ed_{\kv}\ed_{-\kv}f_{\qv}f_{-\qv}\rangle_0=\langle \cos^2\tthk\sin^2\tthq\big(\ud_{\kv}\ud_{-\kv}u_{\qv}u_{-\qv}+ \vd_{\kv}\vd_{-\kv}v_{\qv}v_{-\qv}\big)-\sin 2\tthk\sin 2\tthq \ud_{\kv}\vd_{-\kv}u_{\qv}v_{-\qv}\rangle_0
\label{eq:eeff}
\end{align}
Following Eq.~\ref{eq:secondpertV},
\begin{align}
\frac{1}{N}\sum_{\kv,\qv}\langle \cos^2\tthk\sin^2\tthq\,\ud_{\kv}\ud_{-\kv}u_{\qv}u_{-\qv}\rangle_0&=2\frac{1}{N}\sum_{\kv,\qv}\int_{\omega}\frac{\mathrm{d}\omega}{2\pi}\frac{\delta_{\kv,\qv}\cos^2\tthk\sin^2\tthk}{(i\omega-\omega_{u,\kv})(-i\omega-\omega_{u,-\kv})}\non\\
&=2\frac{1}{N}\sum_{|\kv|<\Lambda}\frac{\delta_{\kv,\qv}\cos^2\tthk\sin^2\tthk}{(\omega_{u,\kv}+\omega_{u,-\kv})}\rightarrow\frac{\sqrt{3}}{8\pi J_1}|\log (h_l-h)|/S
\label{eq:average}
\end{align}
where the factor $2$ at the front of the RHS of Eq.~\ref{eq:average} comes from two choices to contract $\ud u$. Similarly, we obtain other contributions in Eq.~\ref{eq:eeff} and $K$ at $|\log (h_l-h)|/S$ accuracy reads
\begin{align}
K^{(1)}_{\log}=-\sqrt{3}h_0\frac{h_0}{\pi J_1}(\frac{1}{4}+\frac{1}{4}\frac{1}{1+3\alpha}-\frac{1}{2+3\alpha})\frac{|\log (h_l-h)|}{S}.
\label{eq:resK}
\end{align}
Again $\alpha\equiv J_2/J_1-1/3$. A plot of $K^{(1)}_{\log}$ at primary logarithmical order is shown in Fig.~\ref{fig:UUUDlog}.

Following a similar procedure, we obtain $K^{(1)}$ at the secondary logarithmical order $\log S/S$, and reach a similar result, i.e. $K^{(1)}<0$ when $J_2/J_1\neq1/3$. It implies that upon lowering the field even to $h_l-h\gtrsim 1/S$, $K^{(1)}$ remains negative when $J_2/J_1\neq1/3$, i.e. long wave length fluctuations select planar configuration at $J_2/J_1\neq1/3$.

At $J_2/J_1=1/3$, the mixing part $(J_2/J_1-1/3)\Hcal'$ in $\Hcal_{uuud}$ vanishes, and $e,~f,~\phi$ are canonical eigenmodes at $k^2$ order. Thus $K$ at both primary and secondary logarithmical order vanishes. We calculate regular $1/S$ part of $K$ from quantum fluctuations, and define it as $K^{(1)}_b$ (remind that $K^{(1)}_a$ comes from normal ordering). The calculation is straightforward but tedious. First, similar to how Eq.~\ref{eq:QuarticHi} is obtained, a complete $\mc{H}_{i,\kv}$ is obtained by condensing two incoming or two outgoing bosons while keeping all other bosons as non-condensate operators at all $\kv$. The complete expression of $\mc{H}_{i,\kv}$ reads
\begin{align}
\mc{H}_{i,\kv}=&\frac{1}{N}\sum_{\alpha,\kv\in B.Z.}\{\frac{-1}{2}\big[h_0(\cond{\ad_{\beta,0}}\cond{\ad_{\alpha,0}}(a_{\alpha,\kv}a_{\alpha,-\kv}+a_{\beta,\kv}a_{\beta,-\kv})+\xi^{\gamma}_{\kv}\cond{\ad_{\gamma,0}}\cond{\ad_{\gamma,0}}\dd_{\kv}a_{\gamma,\kv}\big]\non\\&
-\frac{1}{2}\xi^{\gamma}_{\kv}\big(\cond{\ad_{\alpha,0}}^2+\cond{\ad_{\beta,0}}^2-4\cond{\ad_{\alpha,0}}\cond{\ad_{\beta,0}}\big)a_{\alpha,\kv}a_{\beta,-\kv}\}+h.c.
\label{eq:quarticHcond}
\end{align}
Second, we express the condensates $\cond{a_{\alpha,0}}$ in terms of $\Delta_1,\Delta_2$ as defined in Eq.~\ref{eq:abcDelta}, express the non-condensed boson in terms of canonical modes $\{\ta,\tb,\tc,\tdd\}$ defined below Eq.~\ref{eq:quadraticH}, and calculate the one-ladder contribution to $K$ from the four magnon branches at all $\kv$. Terms of the form $\cond{V^2\,\tad_{\alpha}\,\tad_{\beta}\, \ta_{\alpha}\,\ta_{\beta}}_{0},~ \cond{V^2\,\tdd\,\tdd \,\td\,\td}_{0}$ contribute to $K^{(1)}_b$, where $\ta_\alpha,\,\ta_\beta$ run over $\{\ta,\tb,\tc\}$. We obtain $K^{(1)}_b=-0.37 J_1/S$. Adding quantum corrections from normal ordering and quantum fluctuations, we have $K^{(1)}=K^{(1)}_a+K^{(1)}_b=\frac{J_1}{S}(\frac{(1+J_2/J_1)}{24}-0.37)=\frac{-0.32J_1}{S}<0$.

\section{D: Evolution of the UUUD state below $h_l$}
From $K<0$ for all $1/8<J_2/J_1<1$ in the limit $S\gg 1$, the relative phase between $\Delta_1,\Delta_2$ is fixed to be zero $\mod \pi$, and Eq.~\ref{eq:QuarticFEapp} can be expressed as:
\begin{align}
E_{\Delta}/N=-\mu (|\Delta_1|^2+|\Delta_2|^2)+\frac{1}{2}(\Gamma-|K|)(|\Delta_1|^2+|\Delta_2|^2)^2.
\end{align}
However, the degeneracy in the ground state manifold is not fully lifted, and $\Delta=\cos\phi\,|\Delta_1|+\sin\phi\,|\Delta_2|$, $\phi\in(0,2\pi)$. As shown in App.~B, the free energy at order $\Delta^6$ breaks the degeneracy and either $\Delta_1$ or $\Delta_2$ is selected depending on the sign in the sixth order term:
\begin{align}
E^{(6)}_C/N&=C\big((|\Delta_1|+i |\Delta_2|)^3+(|\Delta_1|-i| \Delta_2|)^3\big)^2=2C(|\Delta_1|^6 - 15 |\Delta_1|^4 |\Delta_2|^2 + 15 |\Delta_1|^2 |\Delta_2|^4 - |\Delta_2|^6)
\end{align}
If $C>0$, $\Delta_2$ is selected, and if $C<0$, $\Delta_1$ is selected. Obtaining the sign of $C$ requires calculating the six-boson interaction with $1/S$ correction, which requires much heavier numerical calculations than simply comparing the energy between a sequence of coplanar states that can be described by $\Delta=\cos \phi \,|\Delta_1|+\sin \phi\, |\Delta_2|$ close to $\hsa/2$. We find that the difference in their energy can indeed be fitted by the expression of $E^{(6)}_C$, and $C>0$ in this case. Thus $\Delta_2\equiv\cond{f_0}$ is selected. Similar to how the $\Vbar$ state is identified above $h_u$ in App.~A, From Eq.~\ref{eq:RotUUUD}, 
\begin{equation}
\langle a_0\rangle=\frac{-1}{\sqrt{6}}\langle f_0\rangle,~\langle b_0\rangle=\frac{-1}{\sqrt{6}}\langle f_0\rangle,~\langle c_0\rangle=\frac{2}{\sqrt{6}}\langle f_0\rangle,~\langle \dd_0\rangle=0,
\end{equation}
it can be checked that the magnetic order selected by the sixth-order term has of the three up-spins two are tilting in one direction and another in the opposite direction, while the down spin remains intact (see Fig.~\ref{fig:plateauDown}). We also checked that condensates up to $\Delta^3$ order doesn't couple to the $d$-magnon linearly, suggesting that down-spin remains pointing down below $\hsa/2$. If it is the case at even higher deviations from $h_l$, the transition from the coplanar state to the canted stripe state has to be \textit{first order}, which is also consistent with the kink in the energy vs. $h$ plot (Fig.~\ref{fig:Eplot}).

\end{document}